\documentclass[graybox, nosecnum]{svmult}

\usepackage{mathptmx}       
\usepackage{helvet}         
\usepackage{courier}        
\usepackage{type1cm}        
\usepackage{makeidx}         
\usepackage{graphicx}        
\usepackage{cite}
\usepackage{dcolumn}
\usepackage{bm}
\usepackage{epstopdf}
\usepackage{float}
\usepackage{xcolor, colortbl}
\usepackage{multicol}        
\usepackage[bottom]{footmisc}
\usepackage{hyperref}        
\usepackage{soul}            
\usepackage{CJK}
\usepackage{amsmath}        
\usepackage{bm}
\usepackage{animate}
\usepackage{subfigure}

\makeindex             

\begin{document}
\title*{Models for Pairing Phenomena}
\author{Xiang-Xiang Sun and Shan-Gui Zhou \thanks{corresponding author}}
\institute{Xiang-Xiang Sun \at
           School of Nuclear Science and Technology, University of Chinese Academy of Sciences,
           Beijing 100049, China;
           CAS Key Laboratory of Theoretical Physics,
           Institute of Theoretical Physics, Chinese Academy of Sciences,
           Beijing 100190, China;
           \email{sunxiangxiang@ucas.ac.cn}
           \and
           Shan-Gui Zhou \at
           CAS Key Laboratory of Theoretical Physics,
           Institute of Theoretical Physics, Chinese Academy of Sciences,
           Beijing 100190, China;
           School of Physical Sciences, University of Chinese Academy of Sciences,
           Beijing 100049, China;
           \email{sgzhou@itp.ac.cn}
          }
%
%
\maketitle
\abstract{
Pairing effects manifests themselves in many aspects in nuclear systems
ranging from finite nuclei to
nuclear matter and compact stars.
Although with some specific features for nuclear systems,
the mechanism of pairing between nucleons in these systems
resembles that of electrons in superconductors.
The Bardeen-Cooper-Schrieffer (BCS) theory,
the first successful and microscopic theory for superconductivity,
and the Bogoliubov transformation, the generalization of the BCS theory,
have been widely used to describe pairing correlations in nuclear systems.
To deal with the problem of particle number non-conservation
in the BCS method and generalized Bogoliubov transformation,
particle number projection techniques as well as several approaches
which keep the particle number conserved, have been proposed.
In the study of exotic nuclei, which are quantum open systems,
the continuum contributions have to be taken into account.
In this chapter, a thorough but brief discussion of pairing effects
in nuclear systems will be introduced.
Then nuclear models dealing with pairing correlations in nuclear structure properties
will be presented to different extent of details.
Although formulas are given, the emphasis is mainly put
on the basic ideas concerning these models.
}

\section{\textit{Outline}}
\vspace{-0.5cm}
\begin{itemize}
\hypersetup{hidelinks}
\item \hyperref[sec:Introduction]{Effects of nucleon pairing}
\item \hyperref[sec:Mechanism]{The pairing mechanism}
\item \hyperref[sec:Seniority]{The seniority model}
\item \hyperref[sec:BCS]{The BCS theory}
\item \hyperref[sec:Bog]{The generalized Bogoliubov transformation}
\item \hyperref[sec:PNP]{Issues with particle number}
\end{itemize}

\section{\label{sec:Introduction} \textit{1. Effects of nucleon pairing}}

Since the early stage of nuclear physics,
it has been known that an atomic nucleus
is more stable if it consists of even number(s) of protons and/or neutrons,
as is clearly seen both from the fact that there are much more even-even stable isotopes
than odd-mass and odd-odd ones and from the odd-even effects in nuclear masses or binding energies. Such observations indicate that protons (neutrons) like to be coupled into pairs in atomic nuclei.
Indeed, pairing effects have been observed in many essential nuclear phenomena
\cite{Bohr1998_Nucl_Structure_1,Bohr1998_Nucl_Structure_2,Ring1980,Dean2003_RMP75-607,
Brink2005_NuclearSuperfluidity,Broglia2013_50ys-Nuclear-BCS}:
\begin{description}
\item[\textit{The odd-even effect in binding energy}]
The binding energy of an odd-even nucleus is found to be smaller
than the arithmetic mean of binding energies of its two neighboring even-even nuclei.
One can verify this fact
by calculating, e.g., the neutron gaps for even-even nuclei
\begin{equation}
\Delta (N,Z) =-\frac{1}{2} [E_B(Z,\ N-1)+E_B(Z,\ N+1)-2E_B(Z,\ N)],
\end{equation}
with the latest nuclear masses given in AME2020
\cite{Kondev2021_ChinPhysC45-030001,Huang2021_ChinPhysC45-030002,Wang2021_ChinPhysC45-030003}
and can find that they are almost larger than 1 MeV systematically.
$N$ and $Z$ are the number of neutrons and charge number, respectively.
Similar conclusions can also be drawn for the proton gaps.
Note that, besides pairing effects, the odd-even mass staggering
is also related to other mechanisms
\cite{Satula1998_PRL81-3599,Xu1999_PRC60-051301R,Dobaczewski2001_PRC63-024308},
such as the shell structure, single particle properties, and deformations.
\item[\textit{Energy spectra}]
The ground states of all even-even nuclei have the spin-parity $I^\pi =0^+$, which
can be understood as that two nucleons occupying time-reversal
states couple with each other via the pairing interaction.
For an even-even nucleus, the excitation energy of the first
non-collective excited state
can be well estimated as the energy
corresponding to a broken pair \cite{Ring1980}.
The situation in an odd-even nucleus is different:
The ground state of an odd-even nucleus is
almost  completely determined by the last unpaired nucleon
and it can have many excited states in the same energy interval.
\item[\textit{The moment of inertia}]
The moment of inertia of deformed nuclei
can be extracted from the rotational spectra.
The systematics of excitation spectra show that
the moment of inertia from the pure single particle scheme
deviates by a factor of two from the experimental values \cite{Ring1980}.
But when the pairing is included,
theoretical calculations are in good agreement with experimental data.
This is due to the fact that the paired nucleons with the spin zero
contribute little to the rotational angular momentum
\cite{vonOertzen2001_RPP64-1247}.
\item[\textit{The level density}]
In the case of a few nucleons occupying a single $j$-shell,
the energy spectra can be constructed and are all energetically
degenerate corresponding to the various possibilities of angular
momentum coupling.
The number of states per energy unit can easily be estimated and it is found
that the energy spectrum is directly related to the pairing strength.
\item[\textit{Deformations}]
In the well accepted mean-field picture,
if proton or neutron
orbitals below a major shell are all occupied,
the nucleus generally has a spherical shape in its ground state.
However, for open shell nuclei,
single particle levels can be partially filled due to pairing correlations.
The quantum correlations originating from the mix of different
single particle orbitals drive these nuclei to be deformed.
This is reflected by the fact that
there is a shape transition from spherical shapes for closed shell nuclei
to well deformed shapes for nuclei with half-filled shells
\cite{Ring1980}.
\item[\textit{The nuclear fission}]%
The spontaneous fission is a decay mode
in which a nucleus splits into two or more lighter nuclei.
This process requires the system to tunnel under the fission barrier.
The character of the system, whether it is superfluid or not,
affects much the fission dynamics
\cite{Bertsch2013_BCS-50ys-26}.
The occurrence of a systematic difference between even-
and odd-mass nuclei in the fission of,
e.g., U isotopes,
is associated with pairing effects \cite{Bohr1998_Nucl_Structure_2}.
\item[\textit{Halos}]
Most of halo nuclei are close to or located at drip lines
\cite{Tanihata2013_PPNP68-215}.
For nuclei close to drip lines,
the valence nucleons are weakly bound and
pairing correlations provide the possibility to
scatter valence nucleons back and forth in the continuum.
Once the valance nucleons occupy orbitals with low orbital angular momenta,
say, $s$- or $p$-wave, a halo can be formed.
Therefore pairing correlations play a crucial role in the formation of nuclear halos.
The occupation of $s$- or $p$-wave orbitals by weakly bound nucleon(s)
leads to that halo nuclei have a pure neutron (proton) matter
with a very low density surrounding a dense core
\cite{Meng1996_PRL77-3963,Meng2006_PPNP57-470,Meng2015_JPG42-093101},
where pair condensate or strong di-neutron correlations may occur \cite{Hagino2007_PRL99-022506,Sagawa2015_EPJA51-102}.
Thus, halo nuclei may be an ideal prototype to search
for evidence of induced pairing through surface vibrational coupling
\cite{Kanungo2013_BCS-50ys-379}.
The pygmy dipole resonances
\cite{Nakamura2006_PRL96-252502,Kanungo2015_PRL114-192502}
as a results of the oscillation between the core and
the low density neutron matter are also connected with pairing correlations.
Furthermore, halo configurations are related to other
aspects of nuclear structures,
including deformations and the shell evolution,
and the interplay among them results in more
fascinating and complicated phenomena
\cite{Nakamura2014_PRL112-142501,Zhou2017_PoS-INPC2016-373,
Sun2018_PLB785-530,Sun2021_SciBulletin66-2072}.
\item[\textit{Two-nucleon decays}]
Nuclei are unbound with respect to nucleon(s) emission beyond the drip lines.
This feature implies some new radioactivities of which
mostly discussed are one- or two-nucleon decay.
A handful of ground state two-proton emitters have been observed
due to the Coulomb interaction which leads to a Coulomb barrier
hindering the escape of protons from the parent nucleus
\cite{Blank2008_RPP71-046301,Pfuetzner2012_RMP84-567}.
There may be also two-neutron radioactivity and
the only candidate identified so far is $^{26}$O
\cite{Lunderberg2012_PRL108-142503}.
The information of two-nucleon decay is particularly important
for revealing the angular and energy correlations between emitted nucleons
\cite{Michel2010_JPG37-064042},
which are highly correlated to the pairing at the initial stage of the process.
\item[\textit{The backbending}]
In a well deformed superfluid nucleus,
the rotational spectrum follows the $I(I+1)$ law and
the moment of inertia is almost a constant below
a certain critical value of rotational frequency.
In some deformed nuclei,
the observed spectra indicate that
there is a very steep increase of moment of inertia
at high angular momenta.
This is because when a deformed nucleus starts to rotate,
the Coriolis force acts in opposite directions
on the two nucleons of each time-reversal pair.
As a consequence,
the rotation tends to align the spins of the nucleons
by successive breaking of pairs with high angular momentum.
\item[\textit{The neutron-proton pairing}]
For nuclei with $N \approx Z$,
the protons and neutrons near the Fermi levels occupy identical orbitals,
which allows for the appearance of pairs consisting of
a neutron ($n$) and a proton ($p$).
The binding energies show the characteristic $T(T + 1)$
isorotational dependence on isospin $T$,
which means the presence of an isovector pair condensate
that rotates in isospace.
Such rotation implies that the existence of the $np$ condensate
is on an equal footing with the $nn$ and $pp$ condensates
\cite{Frauendorf2014_PPNP78-24}.
\item[\textit{Pairing rotations}]
The energies of $I^\pi=0^+$ states relative to that of a reference nucleus are quadratic as a function of the number of additional pairs.
A typical example is the energies of $0^+$ states of Sn isotopes with the reference nucleus $^{116}$Sn \cite{Broglia2000_PR335-1,Potel2011_PRL107-092501}.
This parabolic behavior is similar to the rotational band in deformed nuclei.
The pairing rotation in atomic nuclei is one kind of Nambu-Goldstone modes
as a sequence of the spontaneous symmetry breaking in the $U(1)$ gauge space,
namely the condensation of Cooper pairs
\cite{Broglia2000_PR335-1,Broglia1973_AP80-60,
Brink2005_NuclearSuperfluidity,Potel2013_RPP76-106301}.
\end{description}

In one word, pairing manifests itself in many aspects of nuclear physics.
Besides the above-mentioned nuclear structure features, pairing is also
related to nuclear astrophysics, such as the thermal evolution
of neutron stars and glitches in pulsar stars
\cite{Dean2003_RMP75-607}.
In nuclear reactions,
a typical example is that
the enhanced two-nucleon transfer cross
section is understood as arising from the collective pairing states
\cite{vonOertzen2001_RPP64-1247}.
This chapter mainly focuses on pairing effects in the study of nuclear structure.

\section{\label{sec:Mechanism}\textit{2. The pairing mechanism}}

\subsection{2.1 The pairing forces}

Atomic nuclei are quantum many-body systems
composed of protons and neutrons.
Apart from relatively weak electric forces,
the interaction between two protons is very similar to that between two neutrons.
The isospin degree of freedom is used to distinguish
proton ($\tau=1/2,\ \tau_z=-1/2$)
and neutron ($\tau=1/2,\ \tau_z=+1/2$) and
a nuclear state can be labeled with
the isospin quantum number $T$, with the third component
$T_z =(N-Z)/2$ for a nucleus with $N$ neutrons and $Z$ protons.
When the relative orbital angular momentum $L$ is even,
two nucleons can couple to $T=0$ (isoscalar) and
$T=1$ (isovector) pairs with the spin $S$ of 1 or 0, respectively.
Thus two neutrons can form a pair with $T=1,\ T_z=1$, and $S=0$ and
two protons with $T=1,\ T_z=-1$, and $S=0$.
A neutron and a proton
can couple to $S=0$ and isospin $T=1$ with $T_z=0$ (isovector)
or $T=0$ (isoscalar) and $S=1$
in order to ensure the antisymmetry of the total wave function of the two nucleons.
The possible types of nucleon-nucleon pairs are listed in Table~\ref{Tab:type_pairing}.

\begin{table}[h]
    \centering
    \caption{Possible types of nucleon-nucleon pairs with isospin ($T$ and $T_z$) and total spin ($S$).}
    \begin{tabular}{|p{1.5cm}|p{1.5cm}|p{1.5cm}|p{1.5cm}|}
    \hline\hline
    type &$T$ & $T_z$ & $S$  \\
    \hline
    $pp$ & $1$ & $-1$  &  0  \\
    $pn$ & $1$ & $0$   &  0  \\
    $nn$ & $1$ & $1$   &  0  \\
    $pn$ & $0$ & $0$   &  1  \\
    \hline\hline
    \end{tabular}
    \label{Tab:type_pairing}
\end{table}

Pairing correlations and the phenomena associated with superfluidity
in nuclear physics directly rely on
the underlying interaction,
i.e., the nucleon-nucleon ($NN$) interaction.
Nowadays, the bare $NN$ interaction,
the interaction of two nucleons in the free space,
can be constructed by using some methods to
reproduce nucleon-nucleon scattering phase shifts
in different partial waves labeled by $^{2S+1}L_J$,
where $S,\ L$, and $J$ represent the total spin,
orbital, and total angular momenta, respectively.
The bare $NN$ interaction is characterized by
a strong repulsive core at short distances.
The pairing gap is mainly determined by the
attractive part of $NN$ interactions.
In the $^1S_0$ channel,
the $NN$ interaction is attractive
when the inter nucleon distance is larger than $\sim0.6$ fm.
In pure neutron matter,
neutrons can form Cooper pairs in the weak-coupling limit
with this attractive interaction.
One can also expect that
the system undergoes a Bose-Einstein condensation
into a single quantum state
in the strong coupling regime \cite{Dean2003_RMP75-607}.
The pairing interactions in other channels are also very important,
especially for the study of dynamical and thermal evolution of neutron stars.
For finite nuclei,
although various approaches starting from realistic nuclear
forces have been developed and produced encouraging results,
there are still some issues to be explored further,
such as the understanding of pairing correlations from bare $NN$ interactions,
the influence of the medium polarization on pairing in finite nuclei,
and impacts from Coulomb and three-nucleon forces
\cite{Duguet2013_BCS-50ys-229}.
Thus it is more straightforward to understand the pairing force in a phenomenological way.

The idea of a pairing interaction was already proposed
in the early developments of the traditional shell model
\cite{Mayer1955},
in which single particles move under a central potential
with a strong spin-orbit interaction.
One of the most important effects of the pairing interaction is that it can
couple two identical nucleons to spin zero.
This can be understood in the case of two nucleons interacting
with a short-range attractive effective interaction in a single $j$-shell.
The simplest example is the $\delta$-interaction with the strength $V_0$,
 \begin{equation}
  V(r_{12}) = -4\pi V_0 \delta(\bm r_1-\bm r_2).
\end{equation}
Two identical nucleons in a shell model orbital with angular
momentum $j$ coupling to a total angular momentum $J$ have a wave function
$| j j J M\rangle$ and the interaction energy is
\begin{equation}
  E_J = \langle j j J M |V(r_{12}) |j j J M\rangle =
  -\frac{2j+1}{2} V_0I(j)
  \left|\left\langle J j 0 \frac{1}{2}
  |j \frac{1}{2}\right\rangle \right|^2,
\end{equation}
where
\begin{equation}
 I(j) = \int R^2_j (r_{12}) r_{12}^2 dr_{12},
\end{equation}
is an integral depending on the radial wave
function $R_j(r_{12})$ of the level $j$, and
$\left\langle J j 0 \frac{1}{2}|j \frac{1}{2}\right\rangle $
is the Clebsch-Gordon coefficient.
The energy of the $J=0$ state can be simplified as
\begin{equation}
  E_0 = -\frac{2j+1}{2}V_0 I(j),
\end{equation}
and the energies of other states can be obtained,
\begin{equation}
  E_2 \sim \frac{1}{4}E_0, \quad
  E_4 \sim \frac{9}{64}E_0, \quad
  E_6 \sim \frac{25}{256}E_0
  ,\ \cdots .
\end{equation}
This result shows that the interaction energy of the $J=0$
state is much smaller than other states
and thus the $J=0$ pair is energetically favored.

Besides this $\delta$ interaction,
many phenomenological pairing forces have been proposed,
such as
a constant pairing force \cite{Ring1980},
zero-range pairing force with or without density-dependence
\cite{Dobaczewski1996_PRC53-2809,Meng1998_NPA635-3},
Gogny force \cite{Decharge1980_PRC21-1568},
and separable pairing force of finite range \cite{Tian2009_PLB676-44}.
They have been widely applied in modern nuclear density functional theories
and can describe nuclear superfluidity successfully
\cite{Bender2003_RMP75-121,Meng2006_PPNP57-470,
Vretenar2005_PR409-101,Robledo2019_JPG46-013001}.
Moreover, to simultaneously describe the density
dependence of the neutron pairing gap for both symmetric
and neutron matter, an isospin
dependence in the effective pairing interaction
has also been developed
\cite{Margueron2007_PRC76-064316,Zhang2010_PRC81-044313}
and applied to study the properties of
finite nuclei \cite{Margueron2008_PRC77-054309,Yamagami2009_PRC80-064301,Chen2015_PRC91-047303}.

As mentioned above,
the long-range attractive part of the bare $NN$ force
can lead to the nuclear superfluidity.
Thus many efforts have been made to incorporate realistic forces,
such as low-momentum $NN$ interactions and
Argonne $v_{14}$ and $v_{18}$, to investigate nuclear structure
by using mean-field methods
\cite{Barranco2004_EPJA21-57,Duguet2008_EPJST156-207,
Hebeler2009_PRC80-044321,Pankratov2011_PRC84-014321}.
It has been shown that energy gaps for semi-magic nuclei
can be reproduced in such kinds of calculations.
It should be noted that the implementation of realistic forces
is technically much more complicated than
phenomenological pairing forces
due to the complexities of bare $NN$ forces.
The microscopic understanding of pairing correlations
starting from the underlying forces is still an open question up to now.

\subsection{2.2 Pairing models}

Various approaches have been developed to understand
the pairing phenomena in nuclear systems.
The seniority model \cite{Racah1942_PR062-438,Mayer1950_PR078-22}
solves the problem of
$N$ particles occupying the single $j$-shell
and can be used to understand why the ground
states of even-even nuclei have spin zero and
the spin of an odd-mass nucleus is the
same as the spin of the last unpaired nucleon.
In 1950s the superconductivity in metals was
understood by that two electrons of
opposite momenta attract each other to form a bound state with zero momentum,
which is known as the Cooper pair
and it behaves like a boson
\cite{Cooper1956_PR0104-1189,Bardeen1957_PR0108-1175}.
This is the basic idea of the BCS theory of superconductivity
\cite{Bardeen1957_PR0108-1175}.
The characteristic of a metallic superconductor is primarily the
large energy gap in the spectrum,
corresponding to the energy required to break a Cooper pair.
Following the suggestions of Bohr, Mottelson, and Pines
\cite{Bohr1958_PR0110-936}, the BCS theory
was used to study atomic nuclei by Belyaev
\cite{Belyaev1959_MFMDVS31-11}.
The BCS wave function can be generalized
by using the Bogoliubov-Valatin transformation
\cite{Bogoljubov1958_NuovoCimento7-794,Valatin1961_PR0122-1012}.
Both the BCS approximation and generalized Bogoliubov transformation
have been widely applied in
various mean-field models nowadays to incorporate pairing correlations.
Since the BCS-type wave function does not keep the conservation of particle number,
to deal with this problem
many methods have been proposed,
such as the Lipkin-Nogami method
\cite{Lipkin1960_AoP9-272,Nogami1964_PR0134-B313},
exact solutions
\cite{Richardson1964_NP52-221,Richardson1964_NP52-253}
and particle number conserved methods
\cite{Zeng1983_NPA405-1}
for the pairing Hamiltonian,
and the particle number projection technique in mean-field models
\cite{Sheikh2000_NPA665-71,Anguiano2001_NPA696-467,Anguiano2002_PLB545-62}.
In the following sections,
the basic formulas of these methods and several relevant topics are introduced.

\section{\label{sec:Seniority}\textit{3. The seniority model}}

The ground state of an even-even nucleus has the spin-parity $I^\pi=0^+$ and
the spin of an odd-even nucleus is the same as that of last unpaired nucleon.
This fact can be understood by using the
seniority model, in which
$N$ nucleons interact through a constant pairing force in a single
$j$-shell with the degeneracy of $2j+1$
\cite{Racah1942_PR062-438}.
Setting single particle energies of the orbtials in the $j$-shell
to be zero, then the pairing Hamiltonian is
\begin{equation}
H=-G\sum_{m,\ m'>0}a_{m'}^\dagger
                   a^\dagger_{ -m'}
                   a_{-m}a_m
 = -G\hat{S}_+\hat{S}_-,
\end{equation}
where $m$ is the projection of $\bm j$ on the $z$-axis
and the pair creation and annihilation operators are defined as
\begin{equation}
  \hat{S}_+ \equiv \sum_{m}s_+^{(m)}
  =\sum_{m=1}^{\Omega}a^\dagger_m a^\dagger_{-m},
  \quad
  \hat{S}_-=\left(\hat{S}_+\right)^\dagger,
\end{equation}
with the number of paired states $\Omega=(2j+1)/2$.
The quasi-spin operator proposed in Ref.~\cite{Kerman1961_AP12-300}
is one of the methods to solve this problem and will be introduced here.

For each substate $m$, let
\begin{equation}
\begin{split}
  s_+^{(m)} & = a_m^\dag a^\dag_{-m}, \\
  s_-^{(m)} & = a_{-m}   a_m, \\
  s_0^{(m)} & =\frac{1}{2} \left(a_m^\dag a_m + a_{-m}^\dag a_{-m}-1\right).
\end{split}
\end{equation}
One can get the commutation relations as:
\begin{equation}
\begin{split}
   \left[s_+^{(m)}, s_-^{(m)} \right] & = 2 s_0^{(m)},\\
   \left[s_0^{(m)}, s_+^{(m)} \right] & = s_+^{(m)},\\
   \left[s_0^{(m)}, s_-^{(m)} \right] & = - s_-^{(m)}.
\end{split}
\end{equation}
It is found that such commutation properties
are the same as those of angular momentum operators.
$s_0^{(m)}$ has two eigenvalues of $-1/2$ and $1/2$
corresponding to whether the pair $(m,\ -m)$ is full or empty.
Thus the operator $\bm s^{(m)}$ is similar to the spin operator
and is called ``quasi-spin". The total spin vector is
\begin{equation}
 \bm S = \sum_{m>0} \bm s^{(m)},
\end{equation}
and its third component is
\begin{equation}
  S_0 =\sum_{m>0} s_0^{(m)} = \frac{1}{2}\sum_{m>0}
  \left(a^\dag_m a_m + a^\dag_{-m} a_{-m} -1 \right) =
  \frac{1}{2}\left(\hat N -\Omega\right),
\end{equation}
with the particle number operator $\hat N$.

The pairing Hamiltonian can be rewritten as
\begin{equation}
  H = -G\left( \bm S^2 -S_0^2 +S_0 \right),
\end{equation}
and the pairing energy reads
\begin{equation}
  E(N,S)=-G\left[ S(S+1)
  -\frac{1}{4}(N-\Omega)^2
  +\frac{1}{2}(N-\Omega) \right].
\end{equation}
When the number of particles is even,
$S=\Omega/2,\ \Omega/2-1,\ \cdots$, and $|N-\Omega|/2$.
When the particle number is odd,
$S=(\Omega-1)/2,\ (\Omega-3)/2,\ \cdots$, $|N-\Omega|/2$.
If one defines $S=(\Omega-\sigma)/2$ with $\sigma$ being the
\textit{seniority} quantum number,
the energy is
\begin{equation}\label{eq:seni1}
  E(N,\sigma)=-\frac{G}{4}\left[ \sigma(\sigma+1)
 -2\sigma(\Omega+1) + 2N(\Omega+1) -N^2\right].
\end{equation}
The \textit{seniority} quantum number represents the number of unpaired particles in $j$-shell.
The corresponding pairing gaps are
$G(2\Omega+1)/2$ for even particle number
and $G\Omega $ for odd particle number.
This simple model shows that
\begin{itemize}
  \item The ground state for an even system has the minimal seniority $\sigma=0$
  and $\sigma=1$ for an odd system;
  \item For even $N$, the first excited state is given by $\sigma=2$
  and the excitation energy is $G\Omega$, which is independent on $N$;
  \item When the particle number is much smaller than
  the degeneracy of single $j$-shell ($N\ll\Omega$),
  the ground state energy increases with $G\Omega$ multiplied by the number of pairs.
  This is related to the pair vibrational spectrum and
  a typical example is neutron pair vibration based on $^{208}$Pb
  \cite{Mottelson1976_RMP48-375}.
\end{itemize}

The nucleus $^{210}$Po can be taken an example.
Its low-lying excited states with $I^\pi =2^+,\ 4^+$, and $6^+$ are
nearly degenerate and the excitation energy of the $2^+$ state is 1.18 MeV.
This can be well explained by the seniority scheme as
two protons occupying the $1h_{9/2}$ orbital,
which gives the excitation energy of the first excited state about 1 MeV
with the pairing gap estimated by the empirical formula
$(13.43\pm 1.38)A^{-0.48\pm0.03}$ MeV
\cite{Changizi2015_NPA940-210}.

In addition, the seniority model is useful for nuclei
with the number of protons (neutrons) close to magic
numbers for which the configurations can be well described by single $j$ components.
There are many states which are the mixture of several
$j$ configurations even in semi-magic nuclei
\cite{Gottardo2012_PRL109-162502,Simpson2014_PRL113-132502}.
To study these nuclei,
the seniority model has been generalized to the case of multi $j$-shell
\cite{Arima1966_PTP36-296,Talmi1971_NPA172-1,Shlomo1972_NPA198-81}.
Recently, the generalized seniority model has been successfully
applied to study various properties of Sn isotopes
\cite{Morales2011_PLB703-606,Jia2016_PRC94-044312,Maheshwari2016_PLB753-122}.

\section{\label{sec:BCS}\textit{4. The BCS model}}

Although the seniority model is successful for the description of the energy gaps
related with the nuclear superfluidity,
it is limited to the study of nuclei close to magic ones.
The BCS model was firstly proposed to study superfluidity in mentals
\cite{Bardeen1957_PR0108-1175}
and later applied to nuclear systems.
This method can be combined easily with the mean-field models as shown in
Refs.~\cite{Reinhard1986_ZPA323-13,Ring1996_PPNP37-193}
and has been widely applied to study nuclei close to
the $\beta$-stability line.
In this section, the BCS theory will be presented.

\subsection{4.1 The BCS approximation}

In fermion systems like atomic nuclei,
the Kramers degeneracy ensures
the existence of pairs of degenerate and
mutually time-reversal
conjugate states $(k,\ \bar k)$.
Nucleons occupying such states could be coupled strongly by a short-range force.
Under the independent particle approximation,
the ground state properties are
described by filling the single particle levels from the bottom up
to the Fermi level.
As will be shown below, in this case
the occupation probability of each single
particle level is either zero or one.
Due to the pairing interaction,
pairs of nucleons can be scattered from the levels below the Fermi
level to those above.
One then has to deal with the occupation probabilities ranging from zero to one.
Mathematically, this could be solved by introducing the concept of quasiparticles.
For stable nuclei, the BCS approximation has turned out to be very useful.

The model Hamiltonian with a single particle term and a residual two-body interaction is
\begin{equation}
   H=\sum_{n_1n_2}
   \varepsilon_{n_1n_2}
   a_{n_1}^\dag a_{n_2} +
   \frac{1}{4}
   \sum_{n_1n_2n_3n_4}
   \bar{v}_{n_1n_2n_3n_4}
   a^{\dag}_{n_1}a_{n_2}^\dag
   a_{n_4} a_{n_3},
\end{equation}
where $\bar{v}_{n_1n_2n_3n_4} = \langle n_1n_2|v|n_3n_4\rangle -\langle n_1n_2|v|n_4n_3\rangle$.
This problem can be solved by using the BCS approximation.
The wave function for an even-even nucleus can be
approximated by the BCS wave function as
\begin{equation}
  |\mathrm{BCS}\rangle = \prod_{k>0} (u_k + v_k a_k^\dag a_{\bar k}^\dag)|0\rangle,
  \label{Eq:BCS_wf}
\end{equation}
where $u_k$ and $v_k$ represent variational parameters.
The product runs over half of the configuration space ($k>0$).
$|0\rangle$ represents the vacuum state of single particles.
$v_k^2$ and $u_k^2$ represent the probability that
a certain pair of states $(k,\ \bar k)$ is occupied and empty,
respectively.
The norm of the state requires
 \begin{equation}
   u_k^2 + v_k^2 = 1.
 \end{equation}
It should be mentioned that generally $u_k$ and $v_k$ are complex and
it is reasonable to choose real $u_k$ and $v_k$ to satisfy the variation principle
\cite{Ring1980}.
For $k>0$, that means $\bar k < 0$, one has
\begin{equation}
  u_{\bar k} := u_k,\qquad v_{\bar k} := -v_k.
\end{equation}
$u_k$ and $v_k$ can be determined in such a way
that the total energy of the system is minimal.
As is seen in Eq. (\ref{Eq:BCS_wf}),
the BCS wave function does not conserve the particle number.
Under the condition that the expectation value of the
particle number operator has the desired value $N$
\begin{equation}
  \langle\mathrm{BCS}|\hat N |\mathrm{BCS}\rangle= 2\sum_{k>0} v_k^2 =N,
\end{equation}
the variational Hamiltonian reads
\begin{equation}\label{Eq:Ham}
  H' = H- \lambda \langle\hat N\rangle.
\end{equation}
$\lambda$ is the chemical potential or Fermi energy,
which represents the increase of the energy
with respect to a change in the particle number,
\begin{equation}
  \lambda = \frac{dE}{dN}.
\end{equation}
The particle-number uncertainty reads
\begin{equation}
  (\Delta N)^2 := \langle \mathrm{BCS}|\hat N^2|\mathrm{BCS}\rangle-N^2 = 4\sum_{k>0}u_k^2 v_k^2.
\end{equation}
The expectation value of $H'$ with respect to the BCS state is
\begin{equation}
  \langle\mathrm{BCS}|H'|\mathrm{BCS}\rangle =
  \sum_{k}\left[
  (\varepsilon_{kk}-\lambda)v_k^2 +
  \frac{1}{2}
  \sum_{k'}\bar{v}_{kk'kk'}v_k^2 v_{k'}^2
  \right]
  +\sum_{kk'}\bar{v}_{k\bar k k'\bar{k'}}
  u_kv_ku_{k'}v_{k'}.
\end{equation}

Under the condition of $v_k^2+u_k^2=1$,
the BCS wave function can be determined after obtaining $v_k$.
The variation condition
\begin{equation}
  \delta \langle\mathrm{BCS}|H'|\mathrm{BCS}\rangle = 0,
\end{equation}
yields
\begin{equation}
  \left[\frac{\partial }{\partial v_k}
  -\frac{v_k}{u_k}\frac{\partial}{\partial u_k} \right]
  \langle\mathrm{BCS}|H'|\mathrm{BCS}\rangle = 0.
\end{equation}
After differentiating, one can get the BCS equation
\begin{equation}\label{eq:BCS}
  2\tilde\varepsilon_ku_kv_k+\Delta_k(v_k^2 -u_k^2)=0, \quad k>0,
\end{equation}
with
\begin{equation}
  \tilde\varepsilon_k=\varepsilon_{kk} +
  \frac{1}{2} \sum_{k'}(\bar{v}_{kk'kk'}+
  \bar{v}_{\bar kk'\bar k kk'})v_{k'}^2-\lambda,
\end{equation}
and
\begin{equation}\label{Eq:BCS1}
  \Delta_k= -\sum_{k'>0}\bar{v}_{k\bar k k'\bar{k'}}u_{k'}v_{ k'}.
\end{equation}

\begin{figure}[ht]
	\centering
	\includegraphics[width=5.5cm]{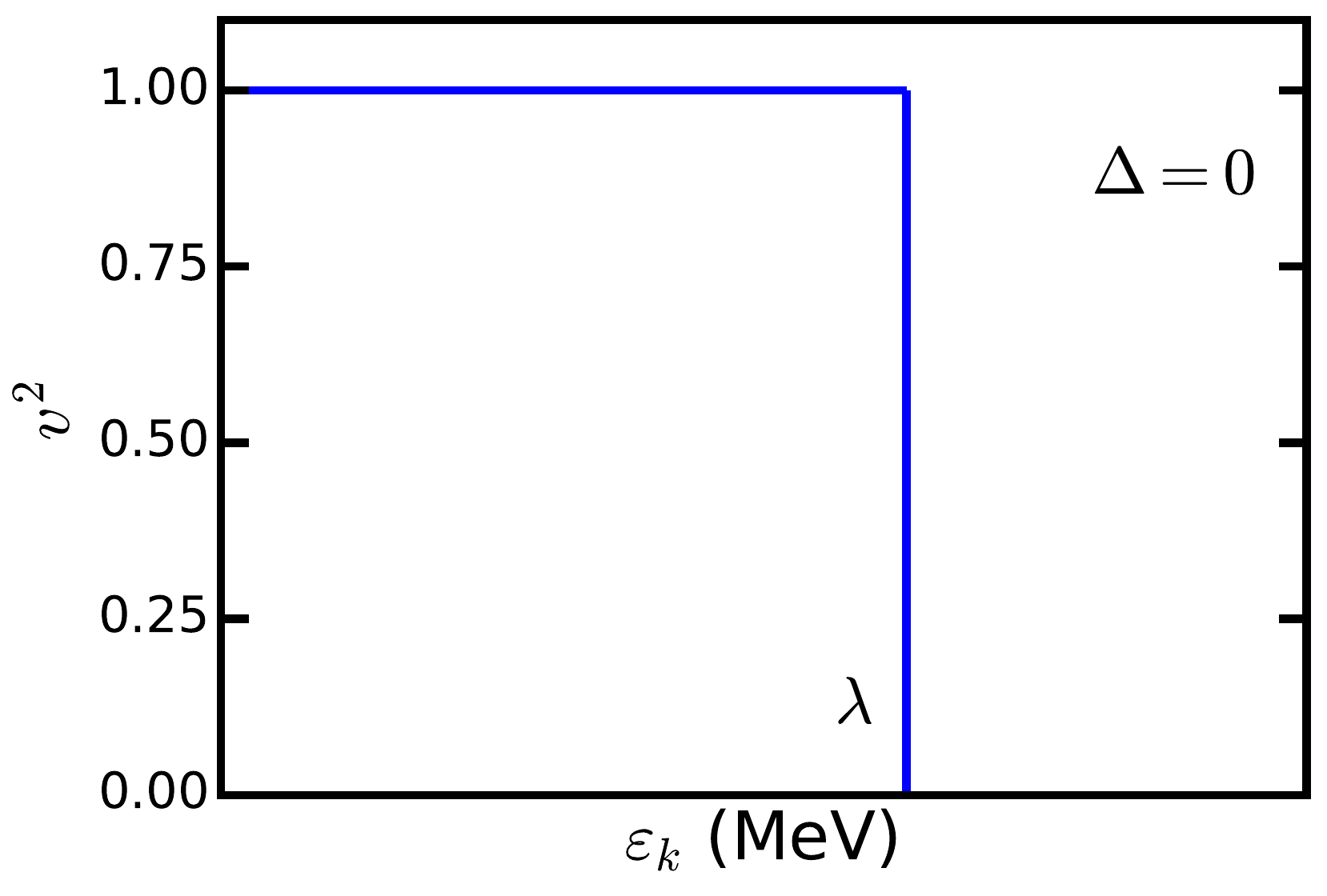}
	\includegraphics[width=5.5cm]{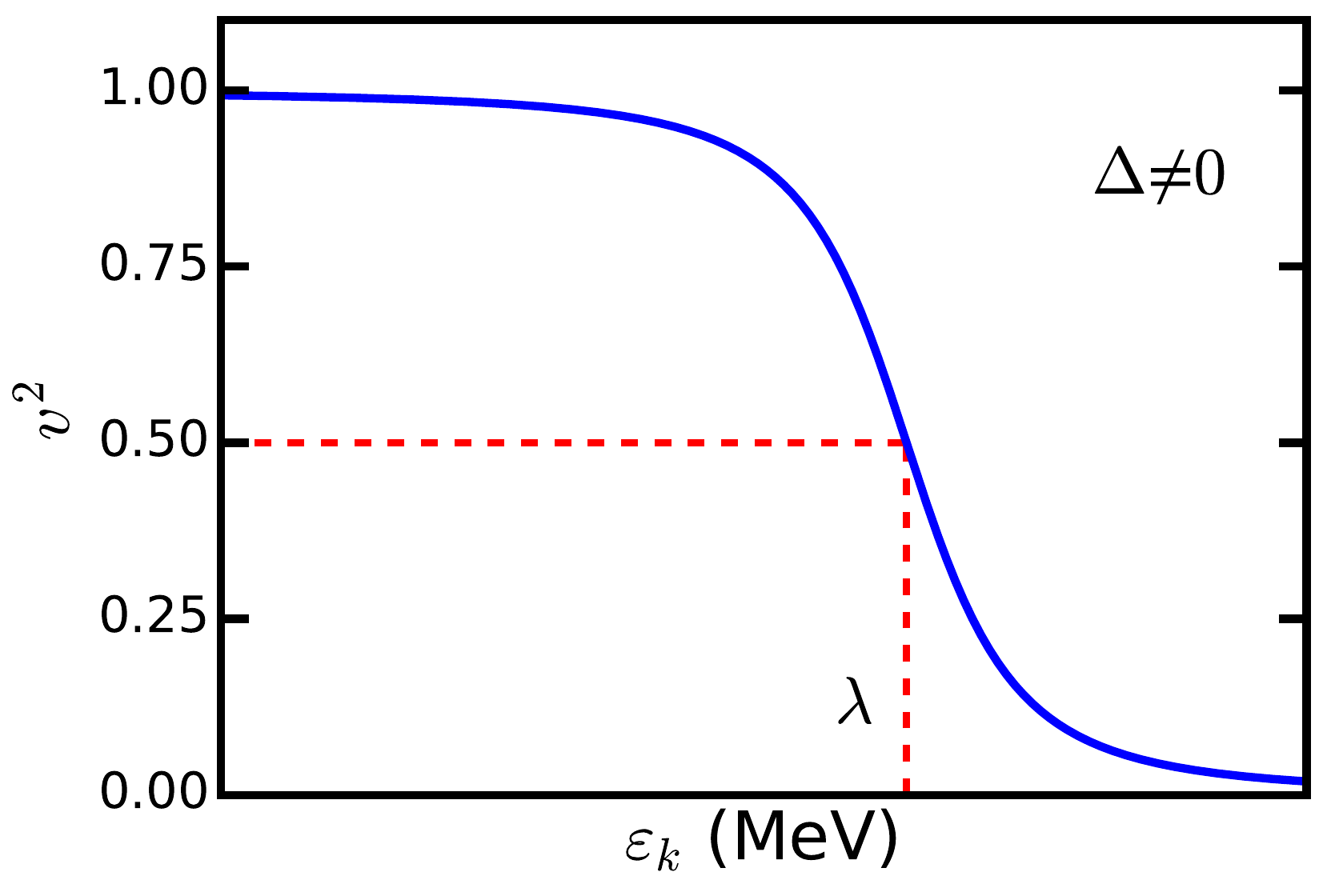}
	\caption{Occupation number as a function of single particle energy for
    pairing gap $\Delta =0$ (left panel) and $\Delta \ne 0$ (right panel).} \label{BCS}
\end{figure}

There is a trivial solution for the BCS equation [cf. Eq.~(\ref{eq:BCS})],
i.e., the pairing gap $\Delta=0$, $v_k^2 =1$ while $u_k^2=0$
for all single particle states below the Fermi level $(\tilde\varepsilon_k \leq \lambda)$.
The non-trivial solution of the BCS equation is
\begin{equation}\label{Eq:BCS2}
  v_k^2 =\frac{1}{2}\left[1 - \frac{\tilde\varepsilon_k}{\sqrt{\tilde\varepsilon_k^2+\Delta_k^2}}\right],
  \quad
  u_k^2 =\frac{1}{2}\left[1 + \frac{\tilde\varepsilon_k}{\sqrt{\tilde\varepsilon_k^2+\Delta_k^2}}\right].
\end{equation}
Then the gap equation can be obtained by inserting
Eq. (\ref{Eq:BCS2}) into Eq. (\ref{Eq:BCS1})
\begin{equation}
 \Delta_k = -\frac{1}{2}\sum_{k'>0}\bar{v}_{k\bar k'k\bar {k'}}\frac{\Delta_{k'}}{\sqrt{\tilde\varepsilon_{k'}+\Delta_{k'}}}.
\end{equation}
In general, these equations are nonlinear and can be solved iteratively.
One can get the occupation probability of each single particle level
determined by the mean-field potential,
shown in Fig. \ref{BCS}.
For the trivial solution,
$v_k^2$ is a step function of the single particle energy.
In the case of finite pairing gap,
the single particle levels around the Fermi energy $\lambda$
are partially occupied and particles can be scattered from below to above the
Fermi energy due to pairing correlations.

The above-mentioned variation procedure usually
leads to the ground state of the system,
which can be compared with the results given by the seniority
model to address the limitation of the BCS method when applying to nuclear systems.
In the case of $N$ particles occupying single $j$-shell,
for the ground state with the seniority zero, one can find
\begin{equation}
  v_k = \sqrt{\frac{N}{2\Omega}},
  \quad
  u_k = \sqrt{1-\frac{N}{2\Omega}}.
\end{equation}
The pairing energy with the BCS model is
\begin{equation}
  E_\mathrm{BCS}^{(N)} = -\frac{1}{2}GN\Omega \left(1-\frac{N}{2\Omega}+\frac{N}{2\Omega^2}\right),
\end{equation}
and this is close to the result obtained from Eq. (\ref{eq:seni1})
when $\Omega \gg N$.
The uncertainty in the particle number is given by
\begin{equation}
  \frac{\Delta N}{ N } = \sqrt{\frac{2}{N}-\frac{N}{\Omega}}.
\end{equation}
Thus the BCS approximation is suitable to study the ground state with a large particle number.
For those systems with small particle number,
the uncertainty from the BCS method is relatively large.

A way to suppress the error due to particle number
fluctuation is the Lipkin-Nogami approximation \cite{Lipkin1960_AoP9-272,Nogami1964_PR0134-B313},
in which the variation Hamiltonian reads
\begin{equation}
H^{\prime\prime}=H-\lambda_1 \langle\hat{N}\rangle -\lambda_2 \langle\hat{N}^2\rangle,
\end{equation}
where $\lambda_1$ represents the Lagrange multiplier
used to constrain the average particle number
and the parameter $\lambda_2$ is determined by
\begin{equation}
\lambda_2=\frac{\left\langle \hat{H} \Delta\hat{N}^2_2\right\rangle}{\left\langle \hat{N}^2 \Delta\hat{N}^2_2\right\rangle}.
\end{equation}
$\hat{N}_2$ is the term of the particle number operator
which projects onto two-quasiparticle states
and $\Delta\hat{N}^2_2 = \hat{N}^2_2- \langle\hat{N}^2_2\rangle$ represents its variance.
In this case,
the Fermi energy is given by $\lambda=\lambda_1+4\lambda_2(N+1)$.
The Lipkin-Nogami method is an approximation to
the particle number projected BCS theory and
has also been applied in modern density functional calculations
\cite{Valor1996_PRC53-172,Reinhard1996_PRC53-2776,
Bender2000_EPJA8-59,Niksic2006_PRC74-064309}
to consider the energy corrections caused by the particle number fluctuation.

The wave function given in Eq. (\ref{Eq:BCS_wf}) corresponds
to the case where all particles are paired.
But for odd-mass or odd-odd nuclei,
the valence particle is unpaired. Therefore,
Eq.~(\ref{Eq:BCS_wf}) cannot be directly used.
One needs to block the valence orbital $k_b$ with
$v_{k_b}^2 =1$ and its time-reversal state is empty.
The corresponding wave function reads \cite{Ring1980,Zhou2001_PRC63-047305}
\begin{equation}
 a_{k_b}^\dag \prod_{k>0,\ k \neq k_b} (u_k + u_k a_k^\dag a_{\bar k}^\dag)|0\rangle.
\end{equation}
Applying this wave function to solve the variation problem,
one can find that
the occupation probabilities for other states are the same as Eq. (\ref{Eq:BCS2}).
The unpaired particle does not contribute to the pairing gap and the total particle number is
$1+2\sum_{k\ne k_b}v_k^2$.
The changes on pairing gaps and occupation probabilities caused by
blocking unpaired particle are called \textit{blocking effects}.
It should be mentioned that this blocking procedure breaks the time-reversal symmetry
and leads to the appearance of currents,
which can be avoided by using the equal filling approximation
\cite{Perez-Martin2008_PRC78-014304,Schunck2010_PRC81-024316}.
This method will be introduced in Sec. \hyperref[hfb_block]{5.3}.

\subsection{4.2 The BCS approximation with resonant states}
\label{subsec:resonant}

\begin{figure}[ht]
	\centering
	\includegraphics[width=5cm]{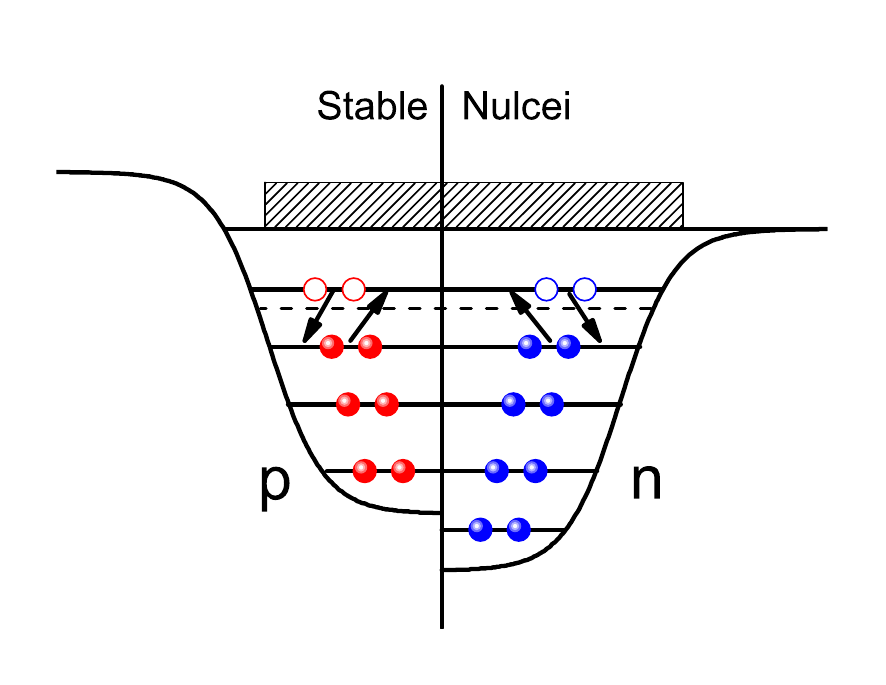}
	\includegraphics[width=5cm]{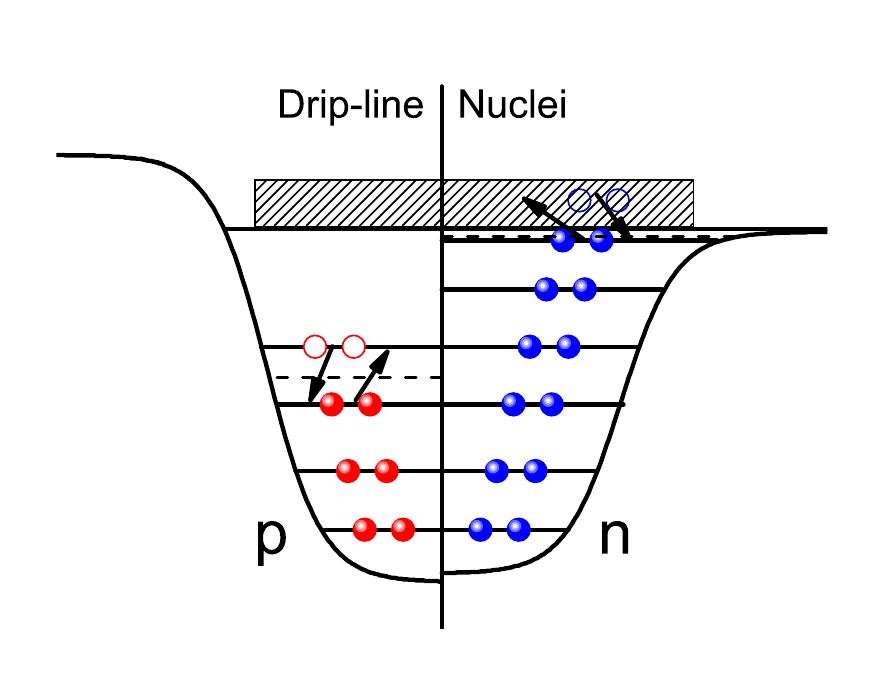}
	\caption{The schematic picture for the difference in pairing
        correlations of stable nuclei (left panel) and drip line nuclei (right panel).
        Normally the last occupied nucleon in stable nuclei has about $8$ MeV binding,
        while it is near the threshold of particle emission in drip line nuclei.
        Pairing correlations in
		exotic nuclei provide the possibility to scatter valence nucleons
		back and forth in the continuum. Taken from Ref.~\cite{Meng2006_PPNP57-470}.
}
 \label{FigC1}
\end{figure}

Nowadays, with the worldwide development of radioactivity-ion-beam facilities,
more and more exotic nuclear phenomena have been observed
\cite{Tanihata2013_PPNP68-215,Nakamura2017_PPNP97-53,
Zhou2017_PoS-INPC2016-373,Yamaguchi2021_PPNP120-103882}.
For a suitable description of exotic nuclei, the contribution from
the continuum should be included
\cite{Dobaczewski2007_PPNP59-432,Michel2010_JPG37-064042,Johnson2020_JPG47-123001}.
However, the conventional BCS
method is only valid for bound states and
not justified for exotic nuclei because it connot
include properly the contribution of continuum states
\cite{Bulgac1980_nucl-th9907088,Dobaczewski1984_NPA422-103}.
For the extremely neutron-rich (or proton-rich) nuclei
near drip lines,
the contribution from the continuum plays an important role as
schematically shown in Fig.~\ref{FigC1}.
The conventional BCS
approach involves unphysical states and the density contributed from continuum is non-local.
For these nuclei, one must either investigate the
detailed properties of continuum states and include the coupling
between the bound state and the continuum by extending the BCS
method to resonant BCS (rBCS) theory
or treat pairing correlations by using the generalized Bogoliubov transformation.
In the following, the way to include the
contribution of resonant states within the rBCS method will be given.

Resonant states are important for determining the pairing properties of the
ground state of nuclei far from the $\beta$-stability line
\cite{Meng2006_PPNP57-470}.
Considering the contribution of continuum
with the BCS approximation can make this method suitable for the study of exotic nuclei.
This idea was first realized in the
Hartree-Fock framework
\cite{Sandulescu2000_PRC61-061301R}.
The main point is that the continuum can be discretized and regarded as
a set of discrete states with the corresponding
level density,
more details can be found in
Refs. \cite{Sandulescu2000_PRC61-061301R,Sandulescu2003_PRC68-054323}.
In the rBCS method,
pairing gaps for bound states are
\begin{equation}\label{eq:gapr1}
\Delta_i
     = \sum_{j}V_{i\overline{i}j\overline{j}} u_j v_j
     + \sum_\nu V_{i\overline{i},\nu\epsilon_\nu\overline{\nu\epsilon_\nu}}
       \int_{I_\nu} g_{\nu}(\epsilon) u_\nu(\epsilon) v_\nu(\epsilon)d\epsilon,
\end{equation}
and averaged pairing gaps for resonances are
\begin{equation}\label{eq:gapr2}
\Delta_\nu
    =\sum_{j} V_{\nu\epsilon_\nu\overline{\nu\epsilon_\nu},j\overline{j}} u_j v_j
         + \sum_{\nu^\prime} V_{\nu\epsilon_\nu\overline{\nu\epsilon_\nu},
           \nu^\prime\epsilon_{\nu^\prime}\overline{\nu^\prime\epsilon_{\nu^\prime}}}
           \int_{I_{\nu^\prime}} g_{\nu^\prime}(\epsilon^\prime)
                 u_{\nu^\prime}(\epsilon^\prime)
                 v_{\nu^\prime}(\epsilon^\prime)
                d\epsilon^\prime,
\end{equation}
where $V$ is the interaction matrix element
and $g_\nu(\epsilon) $ is the total level density,
which takes into account the
variation of the localization of scattering states in the energy
region of a resonance (i.e., the width effect) and goes to a
delta function in the limit of a very small width.
Both bound and resonant states contribute to the total particle number,
which reads
\begin{equation}\label{eq:gapr3}
N = \sum_i v_i^2
   + \sum_\nu \int_{I_\nu} g_{\nu}(\epsilon) v^2_\nu (\epsilon) d\epsilon.
\end{equation}
\begin{figure}[t]
\centering
\includegraphics[width=5.5cm]{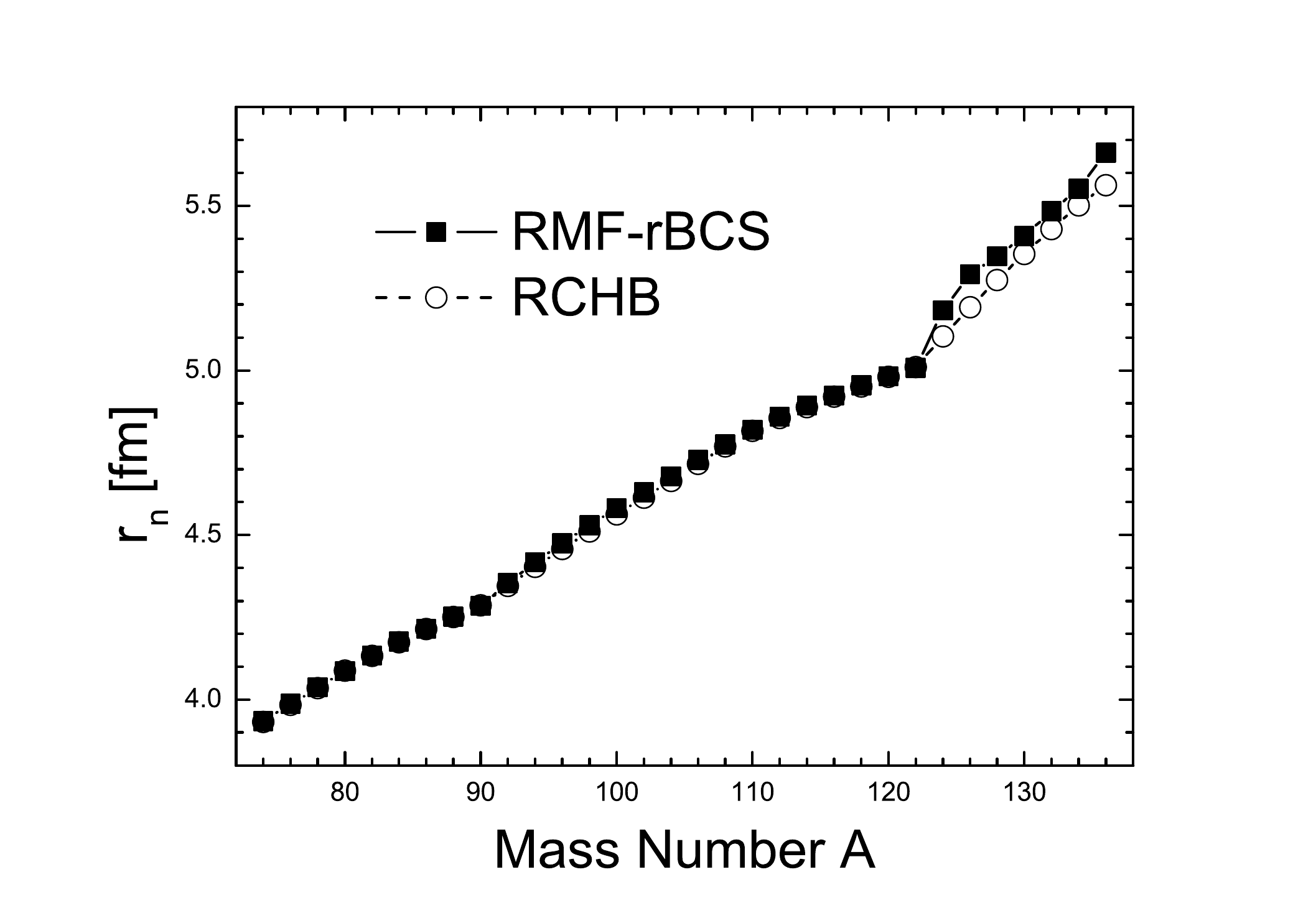}
\includegraphics[width=5.5cm]{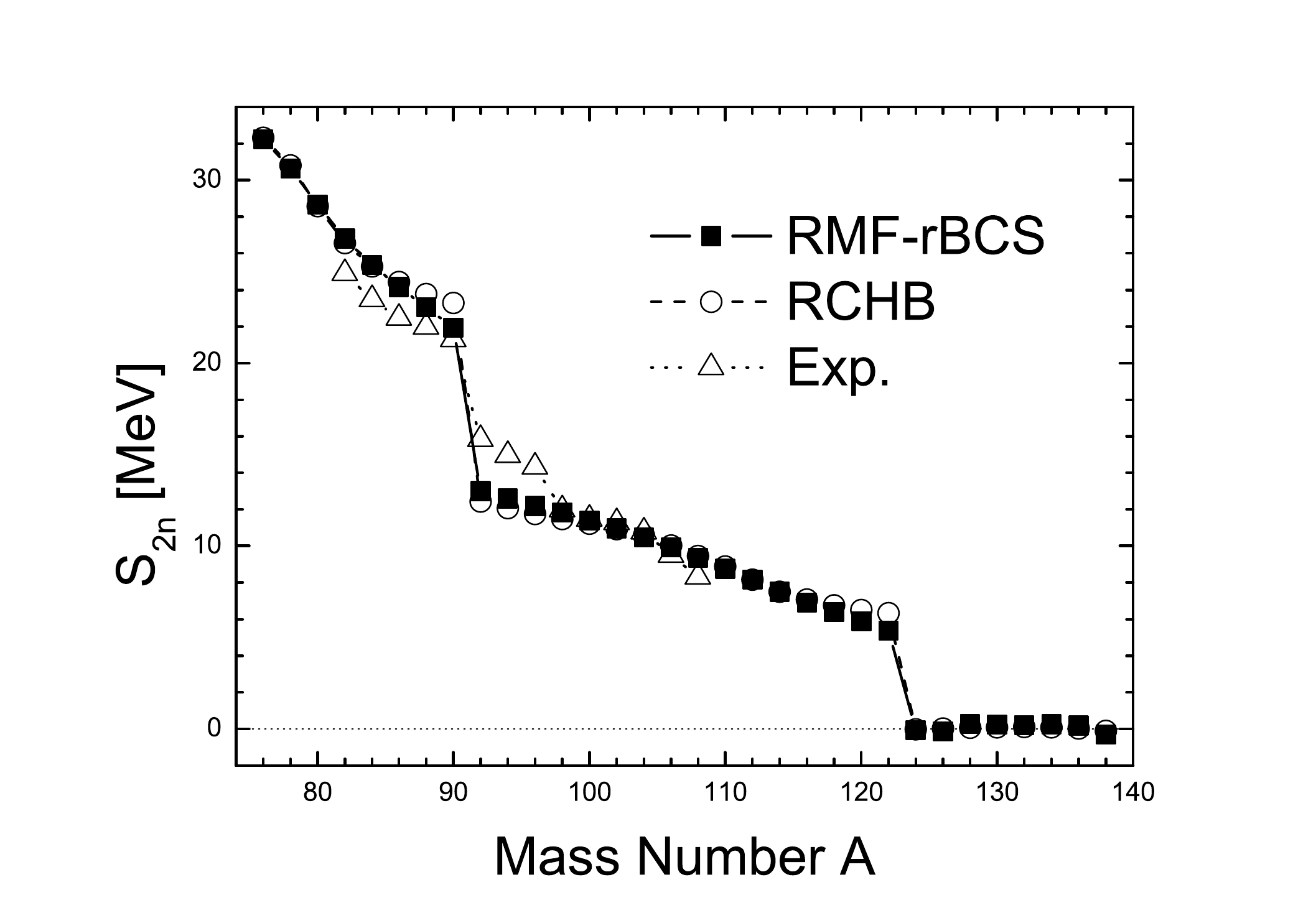}
\caption{The rms neutron radii (left panel) and the two-neutron
separation energies (right panel) of even Zr isotopes as functions
of mass number $A$. Taken from
Ref.~\cite{Sandulescu2003_PRC68-054323}.}\label{FigC8}
\end{figure}

The implementation of the rBCS method in mean-field models
can be achieved straightforwardly by substituting the gap equations
in the conventional BCS method by Eqs.~(\ref{eq:gapr1}--\ref{eq:gapr3}).
The wave functions, energies, and widths of resonant states can be
obtained by using the
scattering-type boundary conditions.
This approximation scheme has been realized in the
relativistic mean field (RMF) model
\cite{Sandulescu2003_PRC68-054323},
called the RMF+rBCS method,
which has been applied to study the neutron-rich Zr isotopes.
It has been shown in Ref.~\cite{Sandulescu2003_PRC68-054323}
that the sudden increase of the neutron radius close to the neutron drip line
depends on a few resonances embedded in the continuum.
As shown in Fig.~\ref{FigC8},
the neutron radii and neutron separation energies from RMF+rBCS calculations
are consistent with the results from the
relativistic continuum Hartree-Bogoliubov (RCHB) calculations,
in which pairing correlations are treated
by using the generalized Bogoliubov transformation
and RCHB equations are solved in coordinate space
\cite{Meng1998_NPA635-3}.

\section{\label{sec:Bog}\textit{5. The generalized Bogoliubov transformation}}

Many properties of nuclei can be
described in terms of independent particle
moving in a mean-field potential \cite{Ring1980}.
In modern nuclear density functional theories,
this mean-field potential can be obtained by
using the Hartree or Hartree-Fock (HF) method.
By combining mean-field models with the BCS method,
the pairing interaction is treated as a residual interaction
and the occupation probabilities are determined
by solving the BCS equations after
obtaining single particle levels.
In other words, the BCS method does not treat the
mean-field and pairing correlations on the same footing.
As a generalization of the HF+BCS method,
the Hartree-Fock-Bogoliubov (HFB) theory
can treat the mean-field and pairing correlations self-consistently.
It has been shown that by formulating
the generalized Bogoliubov transformation
in the coordinate-space representation,
the continuum effects can be properly taken into account
\cite{Bulgac1980_nucl-th9907088,Dobaczewski1984_NPA422-103,
Dobaczewski1996_PRC53-2809,Dobaczewski2013_BCS-50ys-40}.
In this section, the main formulas of the HFB theory will be shown and
some related topics will be given.
For odd-mass or odd-odd nuclei,
the blocking effects must be considered.
Therefore the
blocking method in the HFB theory is also presented.

\subsection{5.1 The Hartree-Fock-Bogoliubov theory}

The HFB wave function $|\Phi\rangle$ is represented as
the vacuum with respect to quasiparticles
\begin{equation}\label{hfb-wf}
  \beta_k |\Phi\rangle = 0, \quad k =1, \dots, M,
\end{equation}
where $\beta_k$ is the quasiparticle annihilation operator
\cite{Ring1980} and $M$ is the number of quasiparticle states.
The HFB wave function can be constructed as
\begin{equation} \label{Eq:gs_even}
  |\Phi\rangle = \prod_k \beta_k |-\rangle,
\end{equation}
with the bare vacuum $|-\rangle$.
The most general linear transformation from the particle operators
$\{a^\dag_l,\ a_l\}$ to the quasiparticle
operators $\{\beta^\dag_k,\ \beta_k\}$ has the form
\begin{equation}
  \begin{pmatrix}
    \beta \\ \beta^\dag
  \end{pmatrix}
  =\begin{pmatrix}
                  U^\dag & V^\dag \\
                  V^T    & U^T
   \end{pmatrix}\
     \begin{pmatrix}
    a \\
    a^\dag
  \end{pmatrix}
  \equiv\mathcal{W}^\dag
  \begin{pmatrix}
    a \\
    a^\dag
  \end{pmatrix}.
  \label{Eq:HFB_matr}
\end{equation}
Since the quasiparticle operators $\{\beta^\dag_k,\ \beta_k\}$
should obey the same fermion commutation relations as the particles,
the matrix $\mathcal{W}$ is unitary,
i.e., $\mathcal{W}\mathcal{W}^\dag=1$.
The corresponding density matrices are defined as
\begin{equation}\label{eq:den_matrix}
\begin{split}
  \rho_{ll'} &   = \langle\Phi| a_{l'}^\dag a_l|\Phi\rangle
                 = \sum_k V_{l'k}V^*_{lk}, \\
  \kappa_{ll'} & = \langle\Phi|a_{l'}a_l|\Phi\rangle
                 = \sum_{k}U_{l'k}V^*_{lk}.
\end{split}
\end{equation}
$\rho$ and $\kappa$ are called the normal and abnormal densities (or density matrix
and pairing tensor), respectively.

A generalized density matrix can be constructed from $\rho$ and $\kappa$
\begin{equation}
  \mathcal{R}=
  \begin{pmatrix}
    \langle\Phi|a_{l'}^\dag a_{l}|\Phi\rangle & \langle\Phi|a_{l'} a_{l}|\Phi\rangle  \\
    \langle\Phi|a_{l'}^\dag a_{l}^\dag|\Phi\rangle  & \langle\Phi|a_{l'}a_{l}^\dag |\Phi\rangle
  \end{pmatrix}
=\begin{pmatrix}
     \rho & \kappa \\
     -\kappa^* & 1-\rho^*
\end{pmatrix}.
\end{equation}
As for the eigenvalue of $\mathcal{R}$, one has
\begin{equation}
  \mathcal{W}^\dag\mathcal{R}\mathcal{W} =
   \begin{pmatrix}
    \langle\Phi|\beta_{l'}^\dag \beta_{l}|\Phi\rangle & \langle\Phi|\beta_{l'} \beta_{l}|\Phi\rangle  \\
    \langle\Phi|\beta_{l'}^\dag \beta_{l}^\dag|\Phi\rangle  & \langle\Phi|\beta_{l'} \beta_{l}^\dag |\Phi\rangle
  \end{pmatrix}
  =\begin{pmatrix}
     0 & 0 \\
     0 & 1
   \end{pmatrix},
\end{equation}
and $\mathcal{R}^2=\mathcal{R}$, which means that
the eigenvalues of $\mathcal{R}$ are $0$ and $1$
with corresponding eigenvectors
$\begin{pmatrix}U\\V \end{pmatrix}$
and
$\begin{pmatrix}V^*\\U^* \end{pmatrix}$.

The expectation value of $H'$ [cf. Eq. (\ref{Eq:Ham})]
with respect to quasiparticle vacuum reads
\begin{equation}\label{Eq:H}
\begin{split}
 \mathcal{E}[\mathcal{R}]
 & = \langle\Phi|H'|\Phi\rangle \\
 & = \sum_{l_2l_2}
     \left( \varepsilon_{l_1l_2} -
            \lambda \delta_{l_1l_2} \right)
     \rho_{l_2l_1}
   + \frac{1}{2}\sum_{l_1l_2l_3l_4}
     \bar v_{l_1l_2l_3l_4}
     \rho_{l_3l_1}\rho_{l_4l_2} \\
   &~~+ \frac{1}{4}\sum_{l_1l_2l_3l_4}
     \bar v_{l_1l_2l_3l_4}
     \kappa_{l_1l_2}^*\kappa_{l_3l_4},
\end{split}
\end{equation}
which is a functional of the general density matrix $\mathcal{R}$.
The variation of the energy functional is
\begin{equation}
  \delta\mathcal{E}
  = \mathcal{E}[\mathcal{R}+\delta\mathcal{R}]
  - \mathcal{E}[\mathcal{R}]
  = \sum_{kk'} \mathcal{H}_{kk'}
    \delta\mathcal{R}_{kk'},
\end{equation}
where the Hamiltonian matrix $\mathcal{H}$ is defined as
\begin{equation}
  \mathcal{H}_{kk'} = \frac{\partial \mathcal{E}[\mathcal{R}]}{\partial \mathcal{R}_{k'k}}.
\end{equation}
Since
\begin{equation}
  \begin{split}
    \frac{\partial \mathcal{E}}{\partial \rho_{k'k}}
    & = \varepsilon_{kk'}-
        \lambda\delta_{kk'} +
        \sum_{pp'}\bar v_{kp'k'p}\rho_{pp'}
    \equiv
    \varepsilon_{kk'} -
    \lambda\delta_{kk'}+ \Gamma_{kk'}, \\
    -\frac{\partial \mathcal{E}}{\partial \kappa^*_{k'k}}
    & = \frac{1}{2}\sum_{pp'}
    \bar v_{kk'pp'}\kappa_{pp'}
    \equiv
    \Delta_{kk'},
  \end{split}
\end{equation}
$\mathcal{H}$ can be expressed explicitly
as
\begin{equation}
  \mathcal{H} = \begin{pmatrix}
                   h & \Delta \\
                  -\Delta^*& -h^* \\
               \end{pmatrix},
\end{equation}
where $h= \varepsilon-\lambda+\Gamma$
is the single particle Hamiltonian
and $\Delta$ is the pairing potential.

One gets the HFB equation
\begin{equation}
  \begin{pmatrix}
                   h & \Delta \\
                  -\Delta^*& -h^* \\
  \end{pmatrix}
  \begin{pmatrix}
    U_k \\
    V_k \\
  \end{pmatrix}
  = E_k
  \begin{pmatrix}
    U_k \\
    V_k \\
  \end{pmatrix},
\end{equation}
where $E_k$ is the quasiparticle energy and
$\left(U_k,\ V_k\right)^T$ are quasiparticle wave functions.
It should be noted that for the study of ground state of an even-even nucleus,
the generalized Bogoliubov transformation
can be reduced to the BCS-transformation
by transforming the quasiparticle basis to the canonical basis
\cite{Ring1980}.

The HFB theory is the basis of modern nuclear density functional theory,
which provides an amazingly successful
description of the complicated many-body system in nuclei all over
the chart of nuclides
\cite{Negele1982_RMP54-913,Ring1996_PPNP37-193,
Bender2003_RMP75-121,Vretenar2005_PR409-101,
Meng2006_PPNP57-470,Jones2015_RMP87-897}.
The HFB theory provides a unified
description of particle-hole ($ph$)
and particle-particle ($pp$) correlations
\cite{Ring1980}
on a mean-field level
by using two average potentials:
The self-consistent Hartree–Fock
field $\Gamma$ which is attributed to long range $ph$-correlations,
and a pairing potential $\Delta$
which corresponds to the $pp$-correlations.
In nuclear density functional theory,
the effective nucleon-nucleon interactions
are constructed from basic symmetries of
the nuclear force and the involved
parameters are determined by
fitting to characteristic experimental data of
finite nuclei and nuclear matter.
These effective interactions are usually adopted in $ph$-channel in
non-relativistic density functional or covariant density functional theories
while another phenomenological interactions are mostly used for $pp$-channel.
It should be noted that in HFB calculation with the Gogny force, the
interactions used in $ph$-channel is a finite range central potential
in the $pp$ channel
\cite{Peru2014_EPJA50-88,Berger2017_EPJA53-214,Robledo2019_JPG46-013001}.

\subsection{5.2 Selected topics}

As mentioned before,
the generalized Bogoliubov transformation has been widely used
in the study of nuclear structure because
it can provide a self-consistent treatment of mean-field and pairing correlations,
which is particularly important for describing the properties of exotic nuclei.
Thus in this part,
several selected topics related to exotic nuclei will be introduced,
including the continuum effects,
the influence of pairing on nuclear size,
and various pairing forces adopted in density functional theory calculations.

\subsubsection{5.2.1 The generalized Bogoliubov transformation and continuum spectra}

The advantage of the HFB theory is that,
based on the quasiparticle transformation,
it unifies the self-consistent
description of single particle orbitals and the BCS pairing theory into a
single variation theory.
By solving the HFB equation,
one can get the quasiparticle energy spectrum,
which contains discrete bound states,
resonances, and non-resonant continuum states
\cite{Dobaczewski2013_BCS-50ys-40}.
For a self-bound system, i.e., $\lambda<0$,
the matter density calculated from quasiparticle wave functions
$V_k$ with their quasiparticle energies being positive is always local
\cite{Dobaczewski1984_NPA422-103}.
Therefore the contribution of continuum effects can be self-consistently
taken into account in the HFB theory.

\begin{figure}
\begin{center}
\includegraphics[width=10cm]{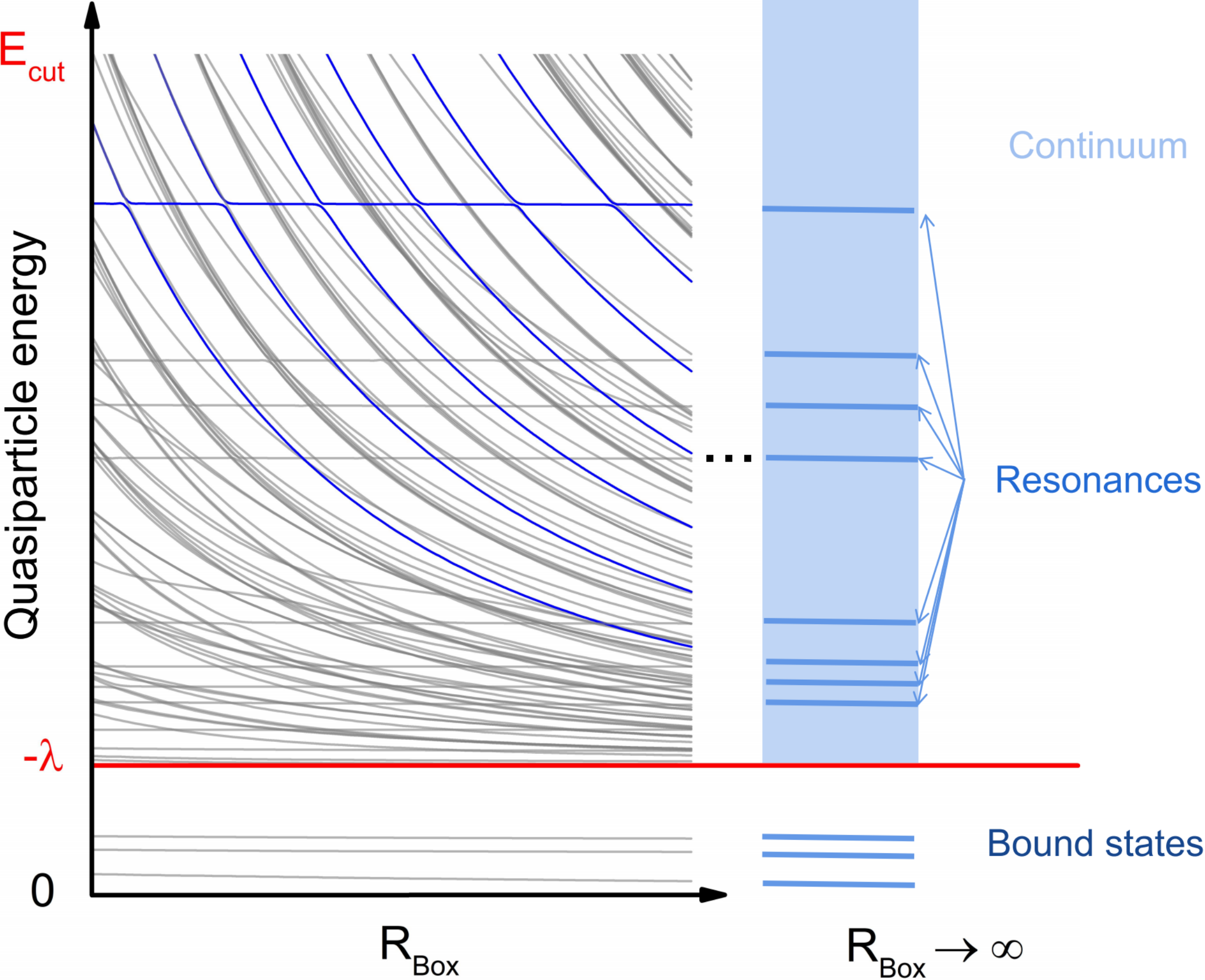}
\caption{Schematic picture of the HFB quasiparticle spectrum.
When solving the HFB equation with the box boundary condition,
the quasiparticle spectrum consists of bound states with quasiparticle
energy smaller than $-\lambda$ and discretized continuum states
with $-\lambda<E_k<E_\mathrm{cut}$.
For each quasiparticle state, the quasiparticle
energy is plotted as a function of the box size.
The blue lines are used to indicate how to find
resonant states by using the stabilization method.
After getting all resonances,
the HFB quasiparticle spectrum is composed of bound states,
continuum resonant states and non-resonant continuum states. }
\label{Fig:continuum}
\end{center}
\end{figure}

The solution of the HFB equations is not so straightforward
because the continuum states must satisfy the scattering boundary conditions.
It has been shown that using the Green's function method
in coordinate space or the Berggren basis,
the ground state from the HFB theory can be obtained and the bound,
resonant, and scattering quasiparticles are well defined
\cite{Belyaev1987_SovJNP45-783,Michel2008_PRC78-044319,
Zhang2011_PRC83-054301,Sun2019_PRC99-054316}.
Alternatively, solving the HFB equation in coordinate space
with the box boundary condition,
the quasiparticle spectrum
consists of bound and discretized continuum states and
the corresponding ground state can be approximately obtained.
It has been shown that by choosing an appropriate box size,
the ground state can be accurately calculated
\cite{Meng1998_NPA635-3,Grasso2001_PRC64-064321,
Pei2011_PRC84-024311,Typel2018_FPhys6-73}.
Besides, several basis expansion methods have also
been proven to be valid for weakly bound nuclei
\cite{Pannert1987_PRL59-2420,Price1987_PRC36-354,
Stoitsov1998_PRC58-2086,Zhou2003_PRC68-034323,
Nakada2018_PRC98-011301R}.
With the box boundary condition in HFB calculations,
the resonant states can be well located by using the
the stabilization method \cite{Pei2011_PRC84-024311}.

A schematic picture of the HFB quasiparticle
spectrum is shown in Fig. \ref{Fig:continuum}.
The bound states locate in the region $0<E_k<-\lambda$.
When solving the HFB equation with the box boundary condition,
the continuum is discretized and can be replaced by
a set of discrete non-resonant states with $-\lambda<E_k<E_\mathrm{cut}$,
where $E_\mathrm{cut}$ is the cutoff energy in the quasiparticle space.
With the increase of the box size,
the level density in continuum becomes more dense.
It should be noted that when studying exotic nuclei
by solving the wave function in a box,
a proper description should be independent on
$E_\mathrm{cut}$ and the box size $R_\mathrm{Box}$.
In the HFB theory in coordinate representation,
the contribution from the non-resonance continuum state
plays an important role
and
can be included by calculating the densities with the lower component
of quasiparticle wave function, which is always local
\cite{Dobaczewski1984_NPA422-103,Dobaczewski1996_PRC53-2809}.
By using the canonical transformation \cite{Ring1980},
one can get the single particle levels in the canonical basis.
It is found that some bound single particle states might correspond
to quasiparticle states with $E_k>-\lambda$ such that the couplings between
single particle states below the Fermi energy and those in the
continuum can be properly treated.
Additionally, the treatment and influence of
the quasiparticle state with energy larger than
$E_\mathrm{cut}$,
including resonant and non-resonant ones,
can be found in Ref. \cite{Pei2011_PRC84-024311}
and are not discussed in here.

Pairing correlations can significantly affect properties
of the nuclei close to drip lines
due to the presence of the vast continuum space
available for pair scattering
\cite{Dobaczewski1996_PRC53-2809}.
A typical example is the effect of pairing on nuclear halos,
in which the continuum induced by pairing correlations changes the asymptotic
behavior of particle density and
the occupation probabilities of single particle states near
the Fermi energy in even-even weakly bound system thus influencing its spatial extension
\cite{Bennaceur2000_PLB496-154,Zhou2010_PRC82-011301R,
Hagino2011_PRC84-011303R,Pei2013_PRC87-051302R,
Chen2014_PRC89-014312,Pei2014_PRC90-024317,
Meng2015_JPG42-093101,Sun2020_NPA1003-122011}.
In addition, the pairing coupling to positive-energy single particle states
also influences the nuclear binding
\cite{Dobaczewski1996_PRC53-2809}.
Particularly, the strong coupling to the continuum
lowers the Fermi energy thus influences the range of bound nuclei and
impacts the limit of the nuclear landscape
\cite{Goriely2009_PRL102-152503,Xia2018_ADNDT121--122-1,Zhang2022_ADNDT144-101488}.

\subsubsection{5.2.2 Phenomenological pairing force: finite range vs. zero range}

Usually, there are two-kind of pairing interactions commonly used in modern
nuclear density functional calculations: zero-range forces with or without a density
dependence \cite{Dobaczewski1996_PRC53-2809,Meng1998_NPA635-3}
and finite range forces.
The latter includes Gogny
\cite{Decharge1980_PRC21-1568,Berger1984_NPA428-23,Gonzalez-Llarena1996_PLB379-13}
and separable pairing forces \cite{Tian2009_PLB676-44}.
The zero-range force is relatively simple for numerical calculations,
but it allows a coupling to the very highly excited states.
Therefore an energy cutoff has to be introduced
and the interaction strength has to be properly renormalized with respect to, e.g., pairing gaps \cite{Bender2000_EPJA8-59}.
The Gogny force has a better treatment for the
coupling to the highly excited states,
but it involves more sophisticated numerical techniques.
The separable pairing force is numerically simpler than the Gogny force
and has also been widely used nowadays.

\begin{figure}
\begin{center}
\includegraphics[width=4.6cm,angle=270 ]{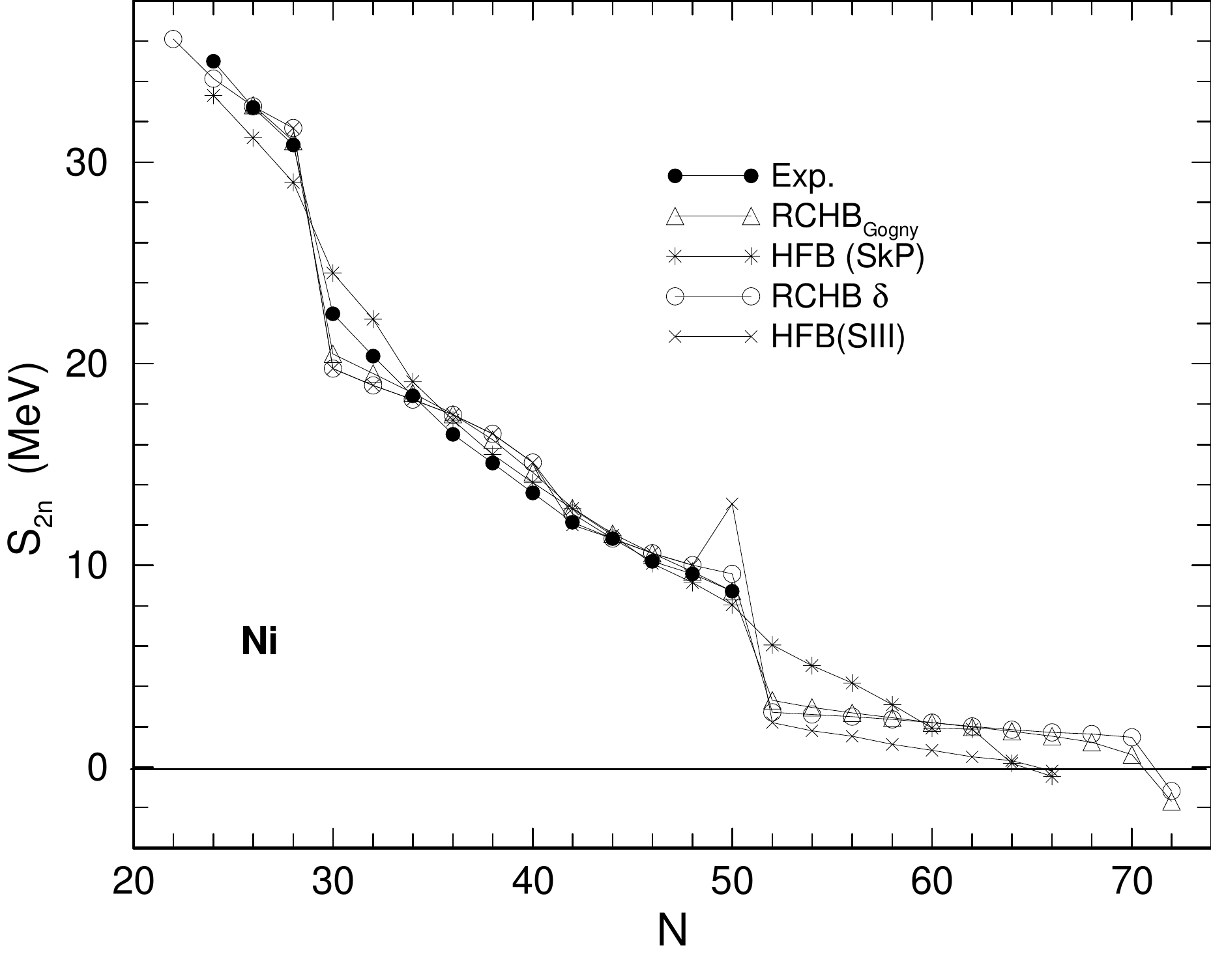}
\includegraphics[width=4.6cm,angle=270]{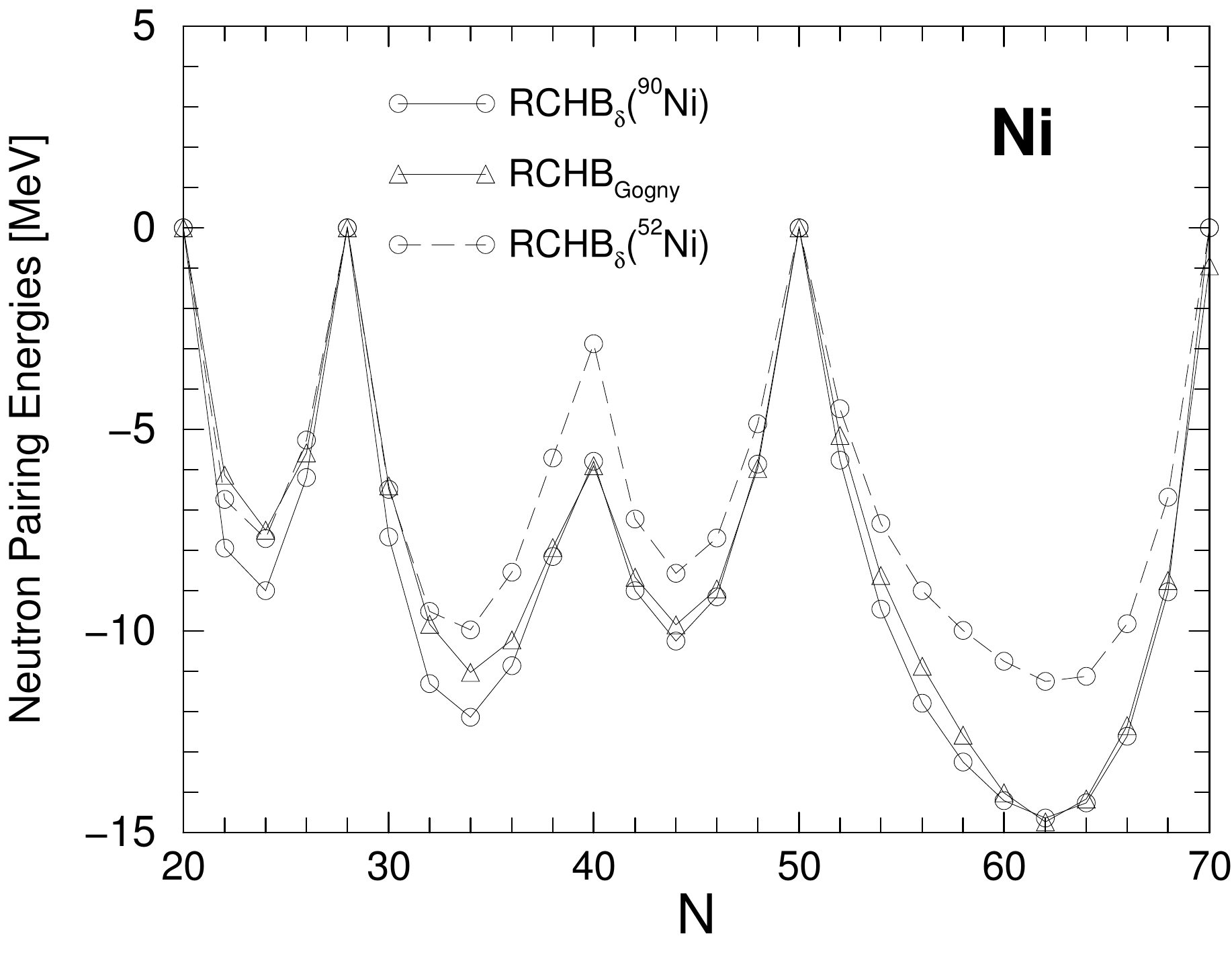}
\caption{Left panel: Two-neutron separation energies $S_{2n}$ of even Ni
isotopes as a function of $N$, including the experimental data
(solid points) and calculation results from the RCHB with $\delta$-force (open circles),
RCHB with
Gogny force (triangles), HFB with SkP interaction (stars), and HFB
with SIII interaction (pluses); Right panel: The neutron pairing
energies for RCHB with Gogny force (triangles) and $\delta$-force
(open circles). The open circles connected by dashed line or solid
line are respectively the pairing energies for RCHB with
$\delta$-force by fitting that of RCHB with Gogny force at
$^{52}$Ni or $^{90}$Ni. Taken from Ref.~\cite{Meng1998_PRC57-1229}.}
\label{FigC10}
\end{center}
\end{figure}

In Ref.~\cite{Meng1998_PRC57-1229}, the proper form of the pairing interaction
is discussed in the framework of the RCHB theory.
The even-even Ni isotopes ranging from the
proton drip line to the neutron drip line are taken as examples.
The pairing correlations are described by using a
density-dependent force with zero range and
the finite range Gogny force and
the results from these two pairing forces were compared.
Through the comparisons of the two-neutron separation
energies $S_{2n}$, the neutron, proton, and matter rms radii,
good agreements have been found between the calculations with
both interactions and the empirical values.
In Fig.~\ref{FigC10}, for example,
the two-neutron separation energies of
Ni isotopes are shown as a function of the neutron number $N$,
including the experimental data (solid points)
and results from the
RCHB with $\delta$-force (open circles),
RCHB with Gogny force (triangles),
and HFB with SkP interaction \cite{Dobaczewski1996_PRC53-2809} (stars)
and SIII interaction \cite{Beiner1975_NPA238-29} (pluses).
The RCHB results with $\delta$-force and Gogny force are almost identical.
There is a strong kink at $N=28$ and a weaker one at $N=50$.
The neutron drip line position is predicted at $^{100}$Ni in both calculations.
The empirical data are known only up to $N=50$.
Comparing with the available empirical data,
the general trend and gradual decline of $S_{2n}$ have been well reproduced.
It is shown clearly in Ref.~\cite{Meng1998_PRC57-1229}
that after proper
renormalization (e.g., fixing the corresponding pairing energies),
observables including two-neutron separation energies and rms radii remain
the same, even when the interaction strength is changed
within a reasonable region.
A further study \cite{Tian2009_PLB676-44}
has shown that the same average gaps
can be given by the Gogny force D1S and $\delta$ force
if the size of the strength is adjusted properly
but the individual matrix elements of the forces
and the matrix elements of the pairing field are very different from each other.
Additionally, the separable approximation is very similar to the full Gogny force.
In conclusion,
for the study of ground state properties
by using mean-filed approaches,
the results from the above-mentioned pairing forces
are almost the same.

\subsubsection{\label{sec:pairing_size}5.2.3 Pairing correlations and nuclear size}

The radius of a nucleus is determined by the spatial distribution of the matter density.
The asymptotic behavior of neutron ground state density from HF calculations
can be approximated as \cite{Bennaceur2000_PLB496-154}
\begin{equation}
   \rho(r) \propto \exp{(-2 \mu r )/ r^2},
\label{densityHF}
\end{equation}
where $\mu = \sqrt{-2m \varepsilon_k} /\hbar$ and $\varepsilon_k$ is the single particle energy
of the least bound orbital with $l=0$. The corresponding mean square radius deduced from the asymptotic solution is
\begin{equation}
   \langle r^2\rangle_{\mathrm{HF}} \propto \frac{\hbar^2}{2m|\varepsilon_k|},
\label{radiusHF}
\end{equation}
which diverges in the limit $\varepsilon_k \rightarrow 0$.
This mechanism, i.e., valence nucleon(s) occupying weakly-bound and low-$l$ orbital, has been used in early interpretations of the nuclear halo phenomenon
\cite{Bertsch1989_PRC39-1154,
Sagawa1992_PLB286-7,Zhu1994_PLB328-1}.

When pairing correlations are included,
the mean square radius deduced from the asymptotic HFB density is
\begin{equation}
   \langle r^2\rangle_{\mathrm{HFB}} \propto \frac{\hbar^2}{2m(E_k-\lambda)}
\label{radiusHFB}
\end{equation}
with the lowest discrete quasiparticle energy
$E_k = \sqrt{(\varepsilon_k-\lambda)^2+\Delta_k^2}$.
When the pairing gap $\Delta_k$ is finite,
the radius does not diverge in the limit of small separation energy
$\varepsilon_k\simeq \lambda \rightarrow 0$.

The asymptotic HF and HFB densities characterized by $l=0$ orbitals were compared in
Ref.~\cite{Bennaceur2000_PLB496-154} and it has been emphasized that pairing
correlations reduce the nuclear size and an extreme halo with infinite
radius cannot be formed in superfluid nuclear systems.
In this sense,
pairing correlations act against the formation of an infinite matter radius.
This is the so called
``pairing anti-halo effect" \cite{Bennaceur2000_PLB496-154}.

In fact, the limiting condition $\varepsilon_k \rightarrow 0$ and the radius deduced from
this $l=0$ orbital alone correspond to an extremely ideal situation, which is
difficult to be found in real nuclei.
In Ref.~\cite{Chen2014_PRC89-014312},
the influences of pairing correlations
on the nuclear size and on the formation of a nuclear halo were
studied in details by using the self-consistent RCHB theory
\cite{Meng1998_NPA635-3}.
It has been shown that pairing correlations not only influence the radius of the orbital, but also
affect the occupation probabilities of the orbitals close to the Fermi level.
As a result, the nuclear radius is dominated by their competition.

\subsection{5.3 Blocking effects}
\label{hfb_block}

To describe odd-mass or odd-odd nuclei,
the blocking effect has to be taken into account.
In this part, the method describing the blocking effect in the HFB appraoch is introduced.
To treat pairing correlations with the generalized Bogoliubov transformation,
the quasiparticle concept is adopted
and the ground state of an even-even nucleus
$|\Phi\rangle$ [cf. Eq. (\ref{Eq:gs_even})] is represented as
a vacuum with respect to quasiparticles \cite{Ring1980}.
For odd-mass and odd-odd nuclei, in practice, the ground state can be
constructed as one quasiparticle state
\begin{eqnarray}
 | \Phi_1 \rangle = \beta^\dag_1 | \Phi_0 \rangle= \beta^\dag_1 \prod_{k} \beta_k | 0\rangle,
\end{eqnarray}
where $\beta_1^\dag$ corresponds to the quasiparticle level to be blocked.
This one quasi-particle state $| \Phi_1 \rangle$ can be regarded as the vacuum
with respect to the set of quasiparticle operators $(\beta'_1,\ \ldots,\ \beta'_M)$ with
\begin{eqnarray}
 \beta'_1 = \beta^\dag_1, \qquad
 \beta'_2 = \beta^{}_2, \qquad \ldots , \qquad
 \beta'_M = \beta^{}_M ,
\end{eqnarray}
and the exchange of the operators $\beta^\dag_1 \leftrightarrow \beta^{}_1$ forms
a new set of quasiparticle operators $(\beta'_1,\ \ldots,\ \beta'_M,\ \beta'^\dag_1,\
\ldots,\ \beta'^\dag_M)$, which corresponds to the exchange of the columns
$( U^{}_{l1},\ V^{}_{l1} ) \longleftrightarrow ( V^*_{l1},\ U^*_{l1} )$
in the matrix $\mathcal{W}$ [cf. Eq.~(\ref{Eq:HFB_matr})].
Therefore, the blocking effect in the odd system can be realized by exchanging the creator
$\beta^\dag_1$ with the corresponding annihilator $\beta^{}_1$ in the quasiparticle space.

Next the procedure will be presented of implementing the blocking in axially deformed nuclei \cite{Li2012_CPL29-042101}.
For a fully paired and axially symmetric deformed system with the time reversal symmetry,
the projection of the total angular momentum on the symmetry axis $\Omega$
is a good quantum number and each single particle state has a degeneracy of two.
The HFB equation can be reduced to half dimension $M/2$ and
decomposed into degenerate blocks with quantum numbers $+\Omega$ or $-\Omega$.
The dimension of the corresponding density and abnormal density matrices is $M$.

For an odd system with the $k_b$-th level blocked in the $+\Omega$ subspace,
the time reversal symmetry is violated and currents appear in the system.
These currents are axially symmetric, i.e., $\Omega$ remains a good quantum number,
but the quasiparticle energies are no longer degenerate for the two subspaces,
because the subspace with $+\Omega$ contains the odd particle and
the subspace with $-\Omega$ contains an empty level.
Therefore, in principle, one has to diagonalize the HFB equation in the whole quasiparticle
space composed of $+\Omega$ and $-\Omega$ subspaces.
With the equal filling approximation
\cite{Perez-Martin2008_PRC78-014304,Schunck2010_PRC81-024316},
which has been shown to be valid in dealing with blocking effects \cite{Bertsch2009_PRC79-034306},
one can average over the two configurations
of a particle in the $+\Omega$
space and in the $-\Omega$ space.
The corresponding currents in two subspaces cancel each other and can be neglected.
In this way one can diagonalize the HFB equation in the $+\Omega$ subspace or the $-\Omega$ subspace
and the resulted fields are time reversal symmetric.
In practice the density matrix $\rho$ and the abnormal
density $\kappa$ in two subspaces
are
\begin{eqnarray}
  \rho & = & \left(V^*  V^T \right)_{M \times M}
         + \frac{1}{2}
           \left( U_{k_b} U^{*T}_{k_b} -  V^*_{k_b} V^T_{k_b} \right) ,
    \label{eq:rhodd} \\
  \kappa & = & \left(V^*  U^T \right)_{M \times M}
         -  \frac{1}{2} \left( U_{k_b} V^{*T}_{k_b} + V^*_{k_b} U^T_{k_b} \right) ,
    \label{eq:kapodd}
\end{eqnarray}
where $V_{k_b}$ and $U_{k_b}$ are column vectors in the matrices $V$ and $U$
corresponding to the blocked level.

\section{\label{sec:PNP}\textit{6. Issues with particle number}}

Both the BCS method and the HFB theory have been widely applied to describe nuclear superfluidity,
but the BCS-type and HFB wave functions do not keep the particle number conserved,
which is connected with the Nambu-Goldstone mode of a broken $U(1)$ phase symmetry
\cite{Broglia2000_PR335-1}.
In addition, the spontaneous breaking of $U(1)$ symmetry leads to a
sharp phase transition, i.e., the pairing energy turns from zero to finite at a critical pairing strength \cite{Ring1980}.
Many efforts have been made to remedy the problem of particle-number violation
in the mean-field model, configuration space, and \textit{ab initio} calculations
(see Ref.~\cite{Sheikh2021_JPG48-123001} and references therein).
The simplest way to treat the uncertainty in the particle number
is the Lipkin-Nogami method mentioned before,
but which only considers the corrections on the energy.
Many exact solutions of the pairing Hamiltonian have also been proposed
\cite{Dukelsky2004_RMP76-643}.
In mean-field models, the restoration of particle number of BCS-type or HFB wave
function can be achieved by using the particle number projection (PNP) method.
In this section, PNP in mean-field models and a particle number
conservation method in the many-body
configuration space are introduced.

\subsection{6.1 Exact solutions for pairing Hamiltonian}

The exact numerical solution of the pairing model has been proposed in the 1960s
by Richardson and Sherman \cite{Richardson1964_NP52-221,Richardson1964_NP52-253}.
This method has been extended to a family of exactly-solvable models, called the
Richardson–Gaudin (RG) models and widely applied in various areas of quantum many-body
systems, such as mesoscopic systems, condensed matter,
quantum optics, cold atomic gases, and atomic nuclei
\cite{Dukelsky2013_50ys-Nuclear-BCS,Dukelsky2004_RMP76-643,Qi2015_PRC92-051304R}.
In this method, one does not need to
diagonalize the pairing Hamiltonian, but instead to solve a set
of non-linear equations, called Richardson's equation,
for parameters in the pairing wave functions.
More details of this method
can be found in Refs.~\cite{Brink2005_NuclearSuperfluidity,Dukelsky2013_50ys-Nuclear-BCS,Dukelsky2004_RMP76-643}.

In this part, a shell-model-like approach (SLAP) for the pairing,
dubbed the particle number conservation (PNC) method
developed in the 1980s \cite{Zeng1983_NPA405-1} will be introduced.
In the PNC method, the pairing Hamiltonian is directly
diagonalized in the many-body configuration space and
it has been shown to be more accurate than
the BCS calculation as compared with the exact
solution \cite{Molique1997_PRC56-1795}.
Furthermore, the blocking effects are
taken into account automatically and both odd-mass and even-even
nuclei can be treated on the same footing.
It has been demonstrated that
the number of configurations with significant contributions to the
low-lying excited states of a nucleus is quite
limited \cite{Wu1989_PRC39-666}.
Consequently the concept of many-body
configuration truncation is introduced instead of the single
particle state truncation used in the BCS or generalized Bogoliubov method.
Extensive studies and discussions on the validity of the truncated
many-body configuration spaces as well as the application of the PNC method can be
found in Refs.~\cite{Wu1991_PRL66-1022,Wu1991_PRC44-2566,
Zeng1994_PRC50-746,Zeng2001_PRC63-024305,Zeng2002_PRC65-044307}.
In particular the presence of the low-lying
seniority $\sigma=0$ solutions, which are usually poorly described by using the standard BCS approximation or HFB theory, has been found to
play a role in the interpretation of the spectra of rotating
nuclei.

In the SLAP for pairing, the model Hamiltonian reads
\begin{equation}
      H=H_{\mathrm{s.p.}}+H_{\mathrm{pair}},
\label{eq:SlapH}
\end{equation}
where
$H_{\mathrm{s.p.}}=\sum_{\nu }\epsilon _{\nu }a_{\nu
}^{\dagger}a_{\nu }$
and the pairing Hamiltonian
$H_{\mathrm{pair}}=-G\sum^{\mu\neq\nu}_{\mu,\nu
>0}a_{\mu }^{\dagger}a_{\bar{\mu}}^{\dagger}a_{\bar{%
\nu}}a_{\nu }$ with $G$ the average strength, $\epsilon _{\nu}$
the single particle energy,
and $\nu $
the notation of the each level.
In the case of axially deformed nuclei,
$\nu \equiv \left(\Omega
,\ \pi\right) $.
$\bar{\nu}$ represents the
time-reversal state of $\nu $.

For an even-even nucleus with the total particle number $N=2n$,
the multi-particle configurations (MPC) used to diagonalize the Hamiltonian
are constructed as the following:

(a) The fully paired configurations with the seniority $\sigma=0$:
\begin{equation}
      |\rho _{1}\bar{\rho _{1}}\cdots \rho _{n}\bar{\rho _{n}}\rangle
      = a_{\rho_{1}}^{\dagger}a_{\bar{\rho}_{1}}^{\dagger}\cdots a_{\rho _{n}}^{\dagger}
        a_{\bar{\rho}_{n}}^{\dagger}\left\vert 0\right\rangle ,
\end{equation}

(b) The configurations with two unpaired particles, i.e., seniority
$\sigma=2$:
\begin{equation}
      |\mu\nu\rho _{1}\bar{\rho _{1}}\cdots
      \rho_{n-1}\bar{\rho}_{n-1}\rangle
         = a_{\mu }^{\dagger}a_{\nu}^{\dagger}a_{\rho_{1}}^{\dagger}
           a_{\bar{\rho}_{1}}^{\dagger}\cdots
           a_{\rho _{n-1}}^{\dagger}
           a_{\bar{\rho}_{n-1}}^{\dagger}\left\vert 0\right\rangle,
\end{equation}
where $\mu$ and $\nu$ denote two unpaired levels.
The MPCs with larger $\sigma$ can also be constructed in this way
\cite{Zeng1983_NPA405-1}.

In realistic calculations, the MPC space has to be truncated.
Only configurations with excitation energies smaller than $E_{c}$ are
used to diagonalize the Hamiltonian (\ref{eq:SlapH}),
where $E_c$ is the cutoff energy. The corresponding nuclear
wave function can be expanded as
\begin{equation}
\begin{split}
      \psi^{\beta } =&\sum_{\rho _{1},\cdots ,
                       \rho _{n}}V_{\rho_{1},\cdots,
                       \rho _{n}}^{\beta }|\rho _{1}\bar{\rho _{1}}\cdots
                       \rho _{n}\bar{\rho _{n}}\rangle \\
                       &
                       +\sum_{\mu ,\nu }\sum_{\rho _{1},\cdots ,
                       \rho_{n-1}}V_{\rho _{1},\cdots ,\rho _{n-1}}^{\beta (\mu \nu )}
                       |\mu\bar{\nu}\rho _{1}\bar{\rho _{1}}\cdots
                       \rho_{n-1}\bar{\rho}_{n-1}\rangle +\cdots,
\end{split}
\end{equation}
where $\beta =0$ (ground state), 1, 2, 3, $\cdots$ (excited states).
The occupation probability of the $i$th-level for state $\beta$ is
\begin{equation}
      n_{i}^{\beta}=\sum_{\rho _{1},\cdots ,\rho _{n-1}}
                \left\vert V_{\rho_{1},\cdots ,\rho _{n-1},i}^{\beta}\right\vert ^{2}
                +\sum_{\mu ,\nu }\sum_{\rho _{1},\cdots,
                       \rho_{n-2}}\left\vert V_{\rho _{1},\cdots ,\rho _{n-2},i}
                       ^{\beta (\mu \nu )}\right\vert ^{2}+\cdots,
                \mathrm{ \ \ \ }
                i=1, 2, 3, \cdots .
\end{equation}%

This SLAP for pairing has been implemented
in the relativistic mean field model
\cite{Meng2006_FPC1-38,Shi2018_PRC97-034317,Xiong2020_PRC101-054305}
and Skyrme Hartree-Fock model
\cite{Liang2015_PRC92-064325,Dai2019_ChinPhysC43-084101}
and it turns out that this hybrid model is valid for both ground state properties
and low-lying spectra. In addition, it has also been applied in several cranking models to study
the rotational properties of ground state bands and low-lying high-$K$ multi-quasiparticle bands
\cite{Fu2013_PRC87-044319,Liang2015_PRC92-064325,
Shi2018_PRC97-034317,Xiong2020_PRC101-054305,Zhang2020_PRC101-054303}.

\subsection{6.2 Particle number projection}

In the HFB theory, the wave functions are the vacua
of the corresponding quasiparticle operators,
which do not represent states with good particle number.
The nonconservation of particle number can be restored by projecting an HFB state
$|\Psi\rangle$ onto a state with good particle number
\cite{Sheikh2000_NPA665-71,Anguiano2001_NPA696-467,Anguiano2002_PLB545-62,
Bender2009_PRC79-044319,Sheikh2002_PRC66-044318}
\begin{equation}
 |\Phi^N\rangle = \hat{P}^N |\Psi\rangle=
 \frac{1}{2\pi}\int_0^{2\pi}\frac{1}{e^{i\varphi N}} e^{i\hat{N}\varphi}|\Psi\rangle,
\end{equation}
with the particle number projection operator
\begin{equation}
 \hat{P}^N = \frac{1}{2\pi}\int_0^{2\pi} e^{i\varphi(\hat{N}-N)}d\varphi.
\end{equation}
Then the projected energy is given by
\begin{equation}
 E^N=  \frac{\langle \Phi^N| \hat{H}|\Phi^N\rangle}{\langle \Phi^N|\Phi^N\rangle}
  = \int_0^{2\pi} d\varphi E[\varphi]
    \mathcal{N}^N(\varphi),
\end{equation}
with
\begin{equation}
 E[\varphi]
 =\frac{\langle \Phi_0| \hat{H}|\Phi_\varphi\rangle}{\langle \Phi_0|\Phi_\varphi\rangle}, \quad
 \mathcal{N}^N(\varphi) =
 \frac{e^{-iN\varphi}}{2\pi}
 \frac{\langle \Phi_0| \Phi_\varphi\rangle}{\langle \Phi^N|\Phi^N\rangle},
\end{equation}
where $|\Phi_\varphi\rangle = e^{i\varphi\hat{N}}|\Psi\rangle$.
It has been shown in Refs.~\cite{Sheikh2000_NPA665-71,Anguiano2001_NPA696-467}
that the energy kernel $E[\varphi]$ is similar to Eq. (\ref{Eq:H}) with modified expressions for the pairing field and the HF potential and can be calculated by the generalized Wick's theorem
\begin{equation}
\begin{split}
  E[\varphi]
    \equiv \frac{\langle \Phi_0| \hat{H}|\Phi_\varphi\rangle}{\langle \Phi_0|\Phi_\varphi\rangle}
    =\sum_{\mu}t_{\mu\mu}\rho_{\mu\mu}^{0\varphi} + \frac{1}{2}\sum_{\mu\nu} \bar v^{\rho\rho}_{\mu\nu\mu\nu}\rho_{\mu\mu}^{0\varphi}\rho_{\nu\nu}^{0\varphi}
   +\frac{1}{4}\sum_{\mu\nu}\bar v^{\kappa\kappa}_{\mu\bar\mu\nu\bar\nu}
   \kappa^{\varphi 0 *}_{\mu\bar\mu}\kappa^{0\varphi }_{\nu\bar\nu},
\end{split}
\end{equation}
where $\bar v^{\rho\rho}$ and $\bar v^{\kappa\kappa}$ denote the effective vertices in the $ph$
and $pp$ channels.
The normal and anomalous transition density matrices are
\begin{equation}
\begin{split}
\rho_{\mu\nu}^{0\varphi} &
= \frac{\langle \Phi_0| a_\nu^\dag a_\mu|\Phi_\varphi\rangle}{\langle \Phi_0|\Phi_\varphi\rangle}, \\
\kappa_{\mu\nu}^{0\varphi}&
= \frac{\langle \Phi_0| a_\nu a_\mu|\Phi_\varphi\rangle}{\langle \Phi_0|\Phi_\varphi\rangle}, \\
\kappa_{\mu\nu}^{ \varphi 0 *}&
= \frac{\langle \Phi_0| a^\dag_\mu a^\dag_\nu|\Phi_\varphi\rangle}{\langle \Phi_0|\Phi_\varphi\rangle}.
\end{split}
\end{equation}
The projected HFB equation read
\begin{equation}
    \begin{pmatrix}
      \epsilon^N +\Gamma^N+\lambda^N & \Delta^N \\
      -\Delta^{N*} & -[\epsilon^N +\Gamma^N+\lambda^N]^*
    \end{pmatrix}
    \begin{pmatrix}
      U^N \\ V^N
    \end{pmatrix} = E
        \begin{pmatrix}
      U^N \\ V^N
    \end{pmatrix}.
\end{equation}
where $\Gamma^N = \frac{\partial E^N}{\partial \rho}$ and $\Delta^N = -\frac{\partial E^N}{\partial \kappa^*}$.
All the quantities keep the good quantum number $N$. Detailed formulas
can be found in Ref.~\cite{Sheikh2002_PRC66-044318}.

The implementation of PNP in the BCS model is relatively simple, which was applied
using different methods shown in
Refs.~\cite{Bayman1960_NP15-33,Dietrich1964_PR135-B22,
Fomenko1970_JPA3-8,Janssen1981_ZPA301-255,Ring1980}.
The PNP technique has been applied in mean-field models based on the HFB or HF+BCS equations.
One can make the projection before or after variation
and the latter is technically much more easier and has been widely used in
beyond-mean-field calculations nowadays.
The mean filed calculations with and without the PNP method have shown that
the uncertainties caused by the particle number fluctuation might be important
for determining the positions of drip lines and the stability of rare isotopes \cite{Schunck2008_PRC78-064305,Schunck2016_RPP79-116301,
An2020_ChinPhysC44-074101,Verriere2021_PRC103-054602}.
More details and related topics on PNP can be found in a recent review
\cite{Sheikh2021_JPG48-123001}.


\section{\label{sec:FR}\textit{Further Reading}}

Pairing effects are directly related to many aspects of nuclear physics.
In this chapter, only the basic picture and methodology of the pairing effects in atomic nuclei
are introduced,
especially focusing on the nuclear structure study.
Various additional topics related to pairing are thoroughly covered in the following textbooks:
\begin{itemize}
\item P. Ring and P. Schuck, The nuclear many-body problem, Springer-Verlag, 1980.
\item W. Greiner and J.A. Maruhn, Nuclear models, Springer-Verlag, 1996.
\item D.M. Brink and R.A. Broglia, Nuclear Superfluidity: Pairing in Finite Systems,
    Cambridge University Press, 2005.
\item D.J. Rowe and J.L. Wood, Fundamentals of nuclear models, World Scientific, 2010.
\item R.A. Broglia and V. Zelevinsky (eds.),
Fifty Years of Nuclear BCS: Pairing in Finite Systems, World Scientific, Singapore, 2013
\end{itemize}

More aspects of pairing effects, underlying mechanism, and more detailed derivations of specific implementations of pairing models, are discussed in the following review articles:
\begin{itemize}
\item R.A. Broglia, J. Terasaki, and N. Giovanardi,
      The Anderson–Goldstone–Nambu mode in finite and in infinite systems,
      Phys. Rep. 335, 1 (2000)
\item W. von Oertzen and A. Vitturi,
      Pairing correlations of nucleons and multi-nucleon transfer between heavy nuclei,
      Rep. Prog. Phys. 64, 1247 (2001)
\item D.J. Dean and M. Hjorth-Jensen,
      Pairing in nuclear systems: From neutron stars to finite nuclei,
      Rev. Mod. Phys. 75, 607 (2003)
\item J. Dukelsky, S. Pittel, and G. Sierra,
      Colloquium: Exactly solvable Richardson-Gaudin models for many-body quantum systems,
      Rev. Mod. Phys. 76, 643 (2004)
\item J. Meng, H. Toki, S.-G. Zhou, S.Q. Zhang, W.H. Long, and L.S. Geng,
      Relativistic continuum Hartree-Bogoliubov theory for ground-state properties of exotic nuclei,
      Prog. Part. Nucl. Phys. 57, 470 (2006)
\item G.C. Strinati, P. Pieri, G. R\"opke, P. Schuck, and M. Urban,
      The BCS–BEC crossover: From ultra-cold Fermi gases to nuclear systems,
      Phys. Rep. 738, 1 (2008)
\item S. Frauendorf and A. Macchiavelli,
      Overview of neutron–proton pairing,
      Prog. Part. Nucl. Phys. 78, 24 (2014)
\item N.Q. Hung, N.D. Dang, and L.G. Moretto,
      Pairing in excited nuclei: A review,
      Rep. Prog. Phys. 82, 056301 (2019)
\end{itemize}		

\section{\textit{Acknowledgements}}

The authors would like to thank Wen-Hui Long, Peng-Wei Zhao, Kai-Yuan Zhang, Shuang-Quan Zhang,
and Zhen-Hua Zhang for their comments and suggestions.
The authors have been partly supported by
the National Key R\&D Program of China (Grant No. 2018YFA0404402),
the National Natural Science Foundation of China (Grants
No. 11525524, No. 12070131001, No. 12047503, No. 11961141004,
No. 11975237, No. 11575189, and No. 11790325),
the Key Research Program of Frontier Sciences of Chinese Academy of Sciences (Grant No. QYZDB-SSWSYS013),
the Strategic Priority Research Program of Chinese Academy of Sciences (Grants No. XDB34010000 and No. XDPB15),
the Inter-Governmental S\&T Cooperation Project between China and Croatia,
and
the IAEA Coordinated Research Project ``F41033''.




\end{document}